\documentclass[usenatbib]{mnras}
\usepackage{amsmath,amsfonts,amssymb}
\usepackage{textcomp,gensymb}
\usepackage[T1]{fontenc}
\usepackage{aecompl}
\usepackage{ulem}
\usepackage[pdftex]{graphicx}
\usepackage{aas_macros}
\usepackage[usenames]{xcolor}
\usepackage{color}
\usepackage{times}
\usepackage{subfig}
\usepackage{stfloats}
\usepackage[utf8]{inputenc}
\usepackage{hyperref}
\usepackage{multirow}
\usepackage{enumerate}

\newcommand{\rev}[1]{\textcolor{black}{#1}}

\newcommand{\sbs}[1]{\ensuremath{_\text{#1}}}  

\newcommand{\total}{\ensuremath{\text{d}}}

\title[Inclined debris discs]{Self-gravity of debris discs can strongly change the outcomes of interactions with inclined planets}

\author[Poblete et al.]{\parbox{\textwidth}{Pedro P. Poblete\thanks{Email: pedro.poblete.rivera@uni-jena.de},
Torsten Löhne,
Tim D. Pearce, 
and 
Antranik A. Sefilian\thanks{Alexander von Humboldt Postdoctoral Fellow}}
\vspace{0.15cm}
\\
Astrophysikalisches Institut, Friedrich-Schiller-Universität Jena, Schillergäßchen 2–3, 07745 Jena, Germany
}

\begin{document}
\date{Accepted ... Received ...}

\pagerange{\pageref{firstpage}--\pageref{lastpage}} \pubyear{2023}

\maketitle

\label{firstpage}

\begin{abstract}
Drastic changes in protoplanets' orbits could occur in \rev{the early stages of planetary systems} through interactions with other planets and their surrounding protoplanetary or debris discs. The resulting planetary system could exhibit orbits with moderate to high eccentricities and/or inclinations, causing planets to perturb one another as well as the disc significantly. The present work studies the evolution of systems composed of an \rev{initially} inclined planet and a debris disc. We perform N-body simulations of a narrow, self-gravitating debris disc and a single interior Neptune-like planet. \rev{We simulate systems with various initial planetary} inclinations, from coplanar to polar configurations considering different separations between the planet and the disc. \rev{We find that except when the planet is initially on a polar orbit,} the planet-disc system tends to reach a quasi-coplanar configuration with low vertical dispersion in the disc. When present, the Zeipel--Kozai--Lidov oscillations induced by the disc \rev{pump the planet's eccentricity and, in turn, affect the disc structure. We also find that the resulting disc morphology in most of the simulations looks very similar in both radial and vertical directions once the simulations are converged.} This contrasts strongly with massless disc simulations, where vertical disc dispersion is set by the initial disc-planet inclination and can be high for initially highly inclined planets. The results suggest caution in interpreting an unseen planet's dynamical history based only on the disc's appearance. 

\end{abstract}



\section{Introduction}
\label{sec:intro}

Debris discs are considered to be the last stage in the planetary formation process. They are remnants of protoplanetary discs after dissipating, leaving mostly solid material behind. This solid material is composed of several species, from micrometric dust to big planetesimals of kilometre size \citep[see, e.g.][for a recent review of debris discs]{Hughes+2018}. The disc components interact both among themselves via two-body scattering, self-gravity, and also with the formed planets in the protoplanetary stage and beyond. Observing a debris disc's structure is thus an important piece of information to unravel the dynamical history of a planetary system and is especially useful because debris discs are easier to detect than planets, especially in the outer regions of systems.

Surveys based on the spectral energy distribution (SED) have shown that debris discs are habitual structures present around main-sequence stars \citep[spectral type A--K,][]{Su+2009,Eiroa+2013, Sibthorpe+2018}. The Asteroid and Kuiper belt are the debris discs in our Solar System. The characterisation of structures and populations in the Kuiper belt is essential to trace the history of Neptune, and every model of the Solar System formation must consider it \citep[see][for a review]{Morbidelli+2020}.   

In the last decade, new high-angular resolution images of debris discs have revealed their complex architecture and diversity. An example of the current resolving capability can be found in \citet{Esposito+2020}, which exhibits a large debris disc sample. The structures present include narrow discs \citep[Fomalhaut's cold outer belt, HR~4796, HD~202628,][]{Kalas+2008,Milli+2017,Faramaz+2019}, extended discs \citep[e.g. HR~8799,][]{Booth+2016}, doubled-ringed discs \citep[e.g. HD~107146,][]{Marino+2018}, warped discs \citep[e.g. $\beta$ Pictoris,][]{Apai+2015,Matra+2019}, among others. Even circumbinary debris discs exist, such as 99~Herculis \citep{Kennedy+2012a}, HD~139006, and HD~13161 \citep{Kennedy+2012b}. Understanding debris disc formation and evolution is necessary to explain or reproduce the observed diversity; planet--disc interactions are thought to be among the most important actors in this process.

Planets can sculpt debris discs via three types of gravitational interactions: (a) secular perturbations, which induce long-term orbit variations \citep[see][]{Murray&Dermott1999}, (b) mean-motion resonances, which can create over-densities in the disc \citep{Ozernoy+2000,Wyatt2003,Kuchner&Holman2003}, and (c) scattering, which can create comet populations \citep{Levison&Duncan1997}, and if the planet is embedded in the disc, it can remove planetesimals from the disc and open gaps \citep[e.g.][]{Friebe+2022}. Planets can also trigger stirring processes in the disc \citep{Mustill&Wyatt2009}, driving changes in the disc morphology and dust grain size distribution via collisional cascades \citep{Krivov+2006,Thebault&Augereau2007,Pan&Schilchting2012}. 
Consequently, \rev{studying planet-debris disc interactions represents an important step forward in the theoretical understanding of debris discs.} 

\rev{Most of the existing planet-debris disc studies} usually assume planets and discs evolve in a coplanar configuration; they assume configurations like our Solar System where the mutual inclination is small. Nevertheless, models of planet-planet scattering have shown that more than half of three-planet interactions can produce a stable system with two of them with a relative inclination greater than $10\degree$ \citep{Marzari&Weidenschilling2002,Chatterjee+2008}. In the observational context, \citet{McArthur+2010} report a relative inclination between two Jupiter-like planets of 30$\degree$ in the $\upsilon$ Andromedae system. In the $\pi$ Men system, \citet{Xuan&Wyatt2020} and \citet{Kunovac2021} report an inclination between a cold Jupiter and a super-Earth with $\geq30\degree$. Finally, for the HAT-P-11 system, \citet{Yee+2018} report an inclination between a Jupiter-like planet and a Neptune-like of $\geq50\degree$. This suggests that \rev{non-zero relative inclinations} in planetary systems are not an exotic feature. Additionally, inclination in debris discs is also observed. In HD~113337 and HD~38529, the relative tilt angle between the planet and the disc is estimated to be in the range $17\degree$-- $32\degree$ and $21\degree$-- $45\degree$ respectively \citep{Xuan+2020}. Therefore, the impact of a non-coplanar planet-disc configuration should be explored.

Indeed, previous works have researched the inclined planet-debris disc interaction. For instance, \citet{Pearce&Wyatt2014} explored the effect of an inclined and eccentric planet on a massive disc. \citet{Mouillet+1997} and \citet{Dawson+2011} explored a warp triggered by a misaligned planet in the $\beta$~Pictoris system, while \citet{Kennedy+2012a} explored the effects of a polar binary on a circumbinary debris disc in the 99~Herculis debris disc system. \rev{More recently, \citet{Farhat+2023} studied the dynamics of a debris disc under the combined action of an inner stellar binary and an external inclined planet, taking HD~106906 as an example.} Nevertheless, these works omitted the effects of disc self-gravity, which could significantly alter the outcome. \rev{We refer the reader to \citet{Sefilian+2021,Sefilian+2023} for a detailed investigation of the role of disc self-gravity in planet--debris disc interactions, although in coplanar configurations.}

\rev{Alternatively, the evolution of mutually inclined massive perturbers and discs has been widely explored in the protoplanetary disc context, particularly in circumbinary setups. Through hydrodynamical simulations and analytical or semi-analytical models, it has been possible to understand the effects of the binaries on discs composed of gas and dust. The long-term evolution in the inclination of the circumbinary discs has been studied in \citet{Lai2014} and \citet{Zanazzi&Lai2018}. In addition, \citet{Martin2014} and \citet{Lubow&Ogilvie2017} developed models by considering the Zeipel--Kozai--Lidov oscillations, which can affect the aspect ratios of circumbinary (and circumstellar) discs strongly.}

In this work, we study the effect of the gravitational perturbations of an inclined planet on a self-gravitating narrow disc for the first time. This could add more information about the evolution of inclined debris discs. The remainder of the paper is laid out as follows. The numerical method and the initial set-up for the N-body simulations are described in Section \ref{sec:methods}. The results are given in Section \ref{sec:results}, and discussed in Section \ref{sec:discussion}. Finally, in Section \ref{sec:conclusion}, we draw our conclusions.  

\section{Methods}
\label{sec:methods}

We performed N-body simulations of inclined debris discs with the N-body code {\sc Rebound} \citep{rebound}. We use mass-bearing particles in the simulation; hence, we consider a self-gravitating disc. The integrator used is {\sc mercurius} \citep{Rein+2019}, which is ideal for studying long evolution times and close encounters among particles simultaneously.

\subsection{Simulation setup}

We explore several setups for different relative tilt angles between the planet and the disc, considering three values of planetary semi-major axis. We consider a central star with 1 $M_{\odot}$ orbited by a super-Neptune planet of mass $10^{-4} M_{\odot}$ ($33.3\ M_{\oplus}$) and initial eccentricity of $e_{\rm pl}=0.01$. The initial values considered for the planet's semi-major axis are $a_{\rm pl} = \{0.3,0.5,0.7\}$ au, and for each value of $a_{\rm pl}$ we incline the planet with respect to the initial disc plane by $i_{\rm pl} = \{0\degree,30\degree,60\degree,90\degree\}$. \rev{Finally, because we are considering practically a circular planet at $t=0$ yrs, we set the argument of periapsis, the longitudes of ascending node, and the true anomaly at their default values, $0\degree$.}

\rev{We assume that the disc lying outside the planetary orbit has a mass of} $5\cdot 10^{-5} M_{\odot}$ ($16.6\ M_{\oplus}$), \rev{which is consistent with estimates of some extrasolar debris-disc masses} \citep{Krivov&Wyatt2021}, giving a disc-planet mass ratio of $\mu = 0.5$. For comparison, we also run equivalent simulations with test (i.e. massless) disc particles. The disc is initialized with 120 equal-mass particles. The disc is narrow, extending from 1.0 au to 1.1 au. The orbital parameters for each particle are chosen following a random uniform distribution in a given range. The initial semi-major axis is chosen between the initial boundaries of the disc. The initial eccentricities are in the range $e_{\rm par } = [0,0.025]$, and the inclinations are in $i_{\rm par } = [0 \degree,1.43\degree]$. The rest of the orbital parameters, i.e. the arguments of periapsis, the longitudes of ascending node, and the true anomalies ($\omega_{\rm par },\Omega_{\rm par },$ and $f_{\rm par }$, respectively) are in the range $0\degree$ to $360\degree$. 

\rev{All of our simulations are evolved for 1~Myr. We note that the secular timescale for a particle orbiting around a Solar mass star at 1~au interacting with a 0.1 $M_J$ planet at 0.3~au is 0.1~Myr \citep[e.g.][]{Murray&Dermott1999}, so running to 1~Myr ensures the simulation runs for sufficient time to capture the evolution. Due to the scalability of our simulations according to Kepler's third law, our results can be scaled to different radial separations, times, and masses. Consequently, our simulation with 1~au in length and a max time of 1~Myr is equivalent to a system that evolves until 1~Gyr with a typical 100~au disc size, considering the same masses and relative planet-disc separations are employed.}



\subsection{Definition of disc parameters}

We mainly focus on the evolution of two parameters in the simulations: inclination and eccentricity. To characterise the disc, we use the angular momentum vector of each particle ($\boldsymbol{ h}_{\mathrm{par}}$) given by $\boldsymbol{h}_{\mathrm{par}} = \boldsymbol{r}_{\mathrm{par}}\times\boldsymbol{v}_{\mathrm{par}}$, where $\boldsymbol{ r}_{\mathrm{par}}$ is the position vector, and $\boldsymbol{v}_{\mathrm{par}}$ is the velocity vector of a particle. Accordingly, taking the component-wise \textit{median}, we define the disc's angular momentum ($\boldsymbol{h}_{\rm d}$) as follows:
\begin{equation}\label{eq:disc_l}
\boldsymbol{h}_{\rm d} = \mathrm{median}\left( \boldsymbol{h}_{\mathrm{par}} \right) .
\end{equation}
Similarly, we can also obtain the characteristic debris eccentricity ($e_{\rm d}$):
\begin{equation}\label{eq:disc_e}
e_{\rm d} = \mathrm{median}\left( e_{\mathrm{par}} \right). 
\end{equation}
The variability or dispersion vector $\boldsymbol{\sigma}$  is computed following the \textit{median absolute deviation} prescription, so that
\begin{equation}
\boldsymbol{\sigma}_{h} = \mathrm{median}\left(\left| \boldsymbol{h}_{\mathrm{par}} -\mathrm{median}\left( \boldsymbol{h}_{\mathrm{par}}\right) \right|\right), 
\end{equation}
and $\sigma_e$ is computed the same way but for $e_{\rm d}$. Doing so will make our disc's parameters less sensitive to outliers, such as ejected particles. Once the angular momentum of the disc has been computed, the instantaneous relative tilt angle between the planet and the disc can be obtained as
\begin{equation}\label{eq:Delta_i}
\Delta i = {\rm cos}^{-1} \left(  \frac{\boldsymbol{h}_{\rm d}\cdot \boldsymbol{h}_{\rm pl}} 
{\| \boldsymbol{h}_{\rm d}\|\cdot \|\boldsymbol{h}_{\rm pl}\|} \right),
\end{equation}
where $\boldsymbol{h}_{\rm pl}$ is the planet's angular moment vector.

\section{Results}
\label{sec:results}

\rev{The results are focused on disc evolution and morphology, planet evolution, and the comparison between the results for massive and massless discs. The disc evolution, focussing on the eccentricity and inclination, is presented in Section \ref{sec:disc_evol}, and the relevance of the disc's self-gravity is highlighted in Section \ref{sec:self-gravity disc}, while the planet's evolution is presented in Section \ref{sec:planet_evol}. Finally, we examine the disc's morphology at the end of our simulation in Section \ref{sec:morphology}. As we shall show below, by the end of our simulations, the majority of the simulated systems will converge to an almost coplanar configuration with low vertical dispersion in the disc, contrary to the test particle simulations.}

\subsection{Disc evolution}
\label{sec:disc_evol}

\rev{Fig.~\ref{fig:disc_res} displays the orientation of the disc's angular momentum (measured with respect to the evolving planet, $\Delta i$; see Eq. \ref{eq:Delta_i}) as well as its eccentricity as a function of time. Results are shown for each of the three considered values of the planetary semi-major axis, each with different initial orbital inclinations (shown in different colours).}

\subsubsection{Inclination evolution}
\label{sec:disc_inc_evol}

\begin{figure}
\centering
\begin{center}
    \includegraphics[width=0.47\textwidth]{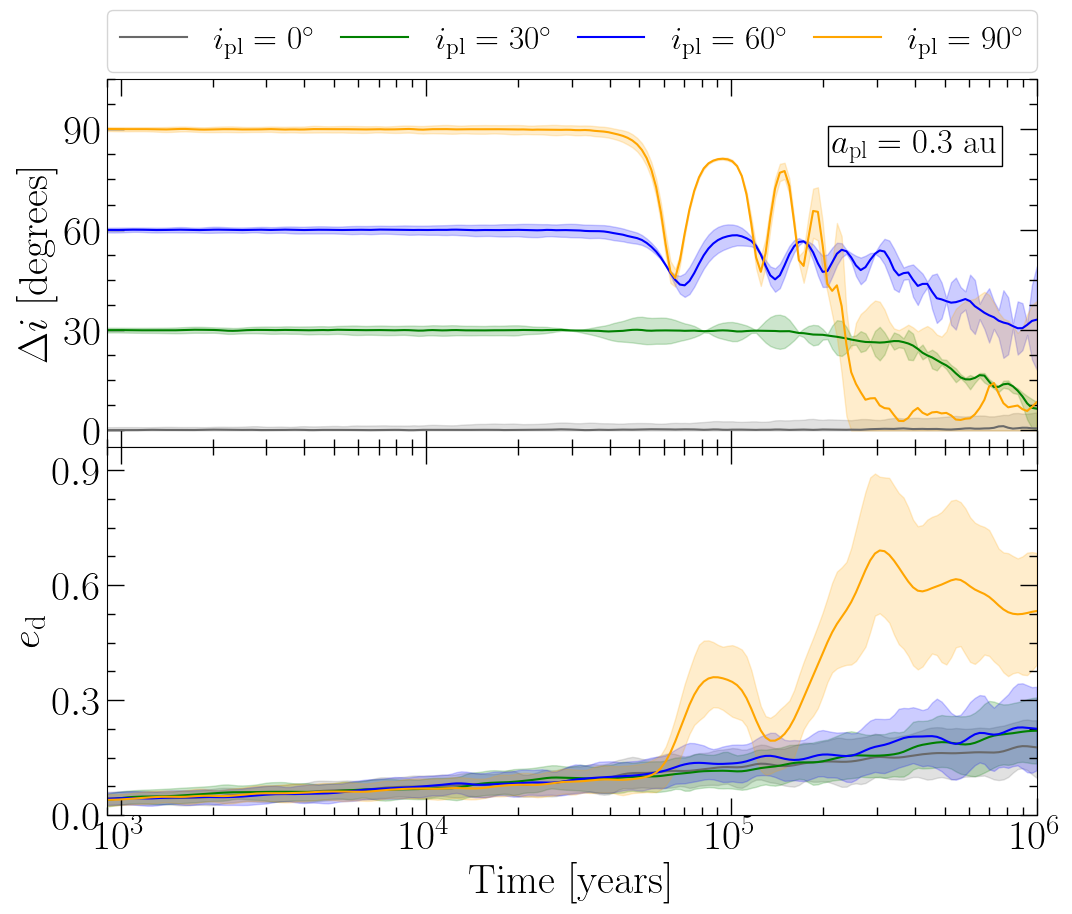}\\
    \includegraphics[width=0.47\textwidth]{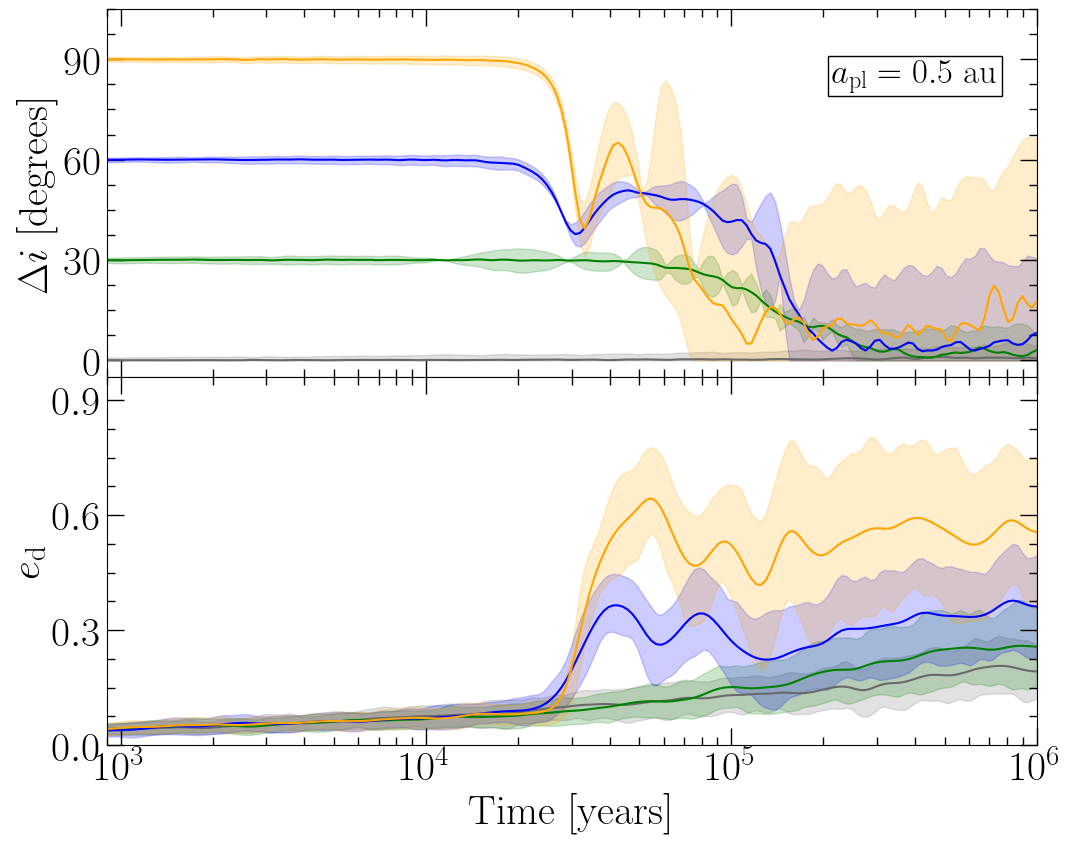}\\
    \includegraphics[width=0.47\textwidth]{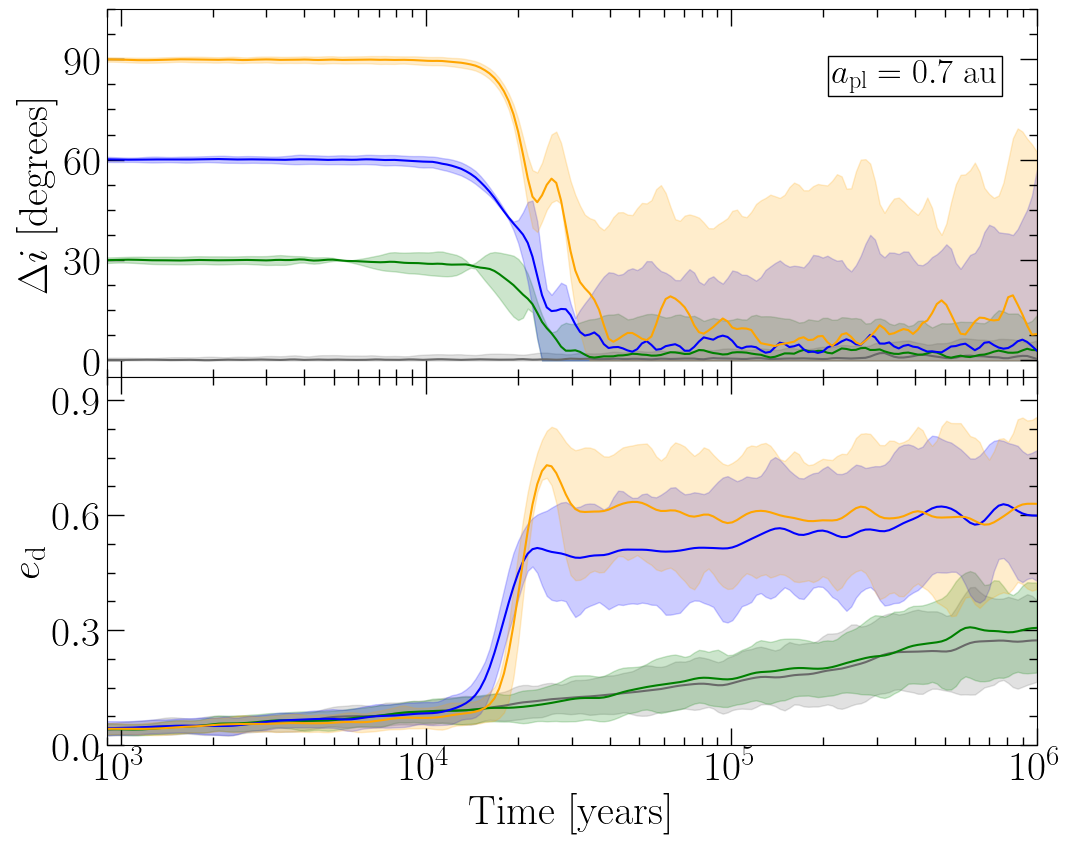} 

    \caption{\rev{The evolution of the disc's angular momentum orientation $\Delta i$, measured relative to that of the evolving planet (Eq. \ref{eq:Delta_i}, top sub-panel), and the median eccentricity of the particles $e_{\rm d}$ (Eq. \ref{eq:disc_e}, bottom sub-panel) for three different values of planetary semi-major axis. The shading corresponds to the spread of the quantities around the median, i.e.,  $\boldsymbol{\sigma}_{h}$ and $\sigma_e$ for the top and bottom sup-panels, respectively. In each panel, the different coloured curves correspond to different initial planet-disc orbit inclinations ($i_{\rm pl}$). It is clear that regardless of the initial conditions, the planet-disc system tends to a quasi-coplanar state, and the disc gains eccentricity; see Section \ref{sec:disc_evol} for further details.}}
    \label{fig:disc_res}
\end{center}
\end{figure}

In the considered evolution time, we generally find that the disc and planet tend towards the same inclination. This is, the planet-disc system evolves until it becomes quasi-coplanar ($\Delta i$ not greater than roughly 10$\degree$, \rev{with a spread in individual debris inclinations around the median}). \rev{This can be seen in each of the panels in Fig. \ref{fig:disc_res}. Note that here we are just exploring the behaviour of the planet-disc mutual inclination $\Delta i$; the planet's inclination relative to the initial disc plane (which also evolves in time) will be discussed in Section \ref{sec:planet_evol}}. The initial polar configurations are the exception because they experience fluctuating $\Delta i$ values greater than 10$\degree$ by the end\footnote{It is worth mentioning that the fluctuations in the median value of $\Delta i$, in the polar case, could be triggered by the low resolution that the simulation has.}, and whilst the $a_{\rm pl} = 0.3$ au and $i_{\rm pl}=60\degree$ simulation (upper panel of Fig. \ref{fig:disc_res}) has not yet become quasi-coplanar by the end of the simulation, its tendency suggests that it will do so if it were evolved further. \rev{Additionally, for the simulations with $a_{\rm pl}=0.3$ au and $a_{\rm pl}=0.5$ au, the transition towards quasi-coplanarity happens fastest in the polar scenario configuration; this is in contrast with the $a_{\rm pl}=0.7$ au simulation, where the transition occurs at similar times for the $i_{\rm pl}=30\degree,60\degree$ and $90\degree$. 
This could suggest an additional, unresolved trend for high inclinations.} Therefore, we can consider that all the non-polar simulations evolve to reach the quasi-coplanar configuration.


Figure \ref{fig:disc_res} also shows that the dispersion in $\Delta i$ correlates with the planet's initial inclination and semi-major axis. For initially larger planetary semi-major axes, simulations with an initially higher $i_{\rm pl}$ exhibit an abrupt increment in their $\Delta i$ dispersion as $\Delta i$ decreases. The dispersion grows until the planet-disc system reaches quasi-coplanarity. The polar case reaches the largest dispersion and a \rev{fluctuating} final median $\Delta i$ value \rev{even for a more distant perturber ($a_{\rm pl}=0.3$ au)}. \rev{This renders the polar case the noisiest simulation in terms of $\Delta i$. This will be further discussed in Section \ref{sec:morphology}.} Nevertheless, in all our runs with massive discs, it is evident that the resulting vertical dispersion is significantly lower than the initial planetary inclination, which would be the value expected for massless discs \citep[e.g.][]{Wyatt+1999,Pearce&Wyatt2014}. \rev{The comparison between self-gravitating and massless discs will be discussed in Section \ref{sec:self-gravity disc}.} 

Our results thus far share a few similarities with those corresponding to planets embedded in protoplanetary discs. For instance, \citet{Xiang-Gruess&Papaloizou2013} studied the evolution of systems similar to ours but for a gaseous disc without self-gravity. They modelled a circular and an eccentric planet with different inclinations. Our results agree with theirs qualitatively, in that the mutual inclinations decay and approach zero over time, with the timescales being shorter for lower inclinations. However, the exact mechanism may differ as their embedded planet experiences dynamical friction due to planetesimal scattering while, e.g., our more distant planet does not. In addition, self-gravity can lead to additional angular momentum transfer within our discs, which is different from the transfer due to the viscous gas in the study of \citet{Xiang-Gruess&Papaloizou2013}.
It is also worth mentioning that their evolution time was significantly shorter than ours because of their assumption of an embedded (i.\,e. much closer) planet in combination with both a higher planet and disc mass. 

\subsubsection{Eccentricity evolution}
\label{sec:disc_ecc_evol}

We now focus on the evolution of the debris' eccentricity. Observing each eccentricity panel in Fig. \ref{fig:disc_res}, we see that all the simulations have an increase in their debris eccentricity value. As for the inclination, the variations in the characteristic debris eccentricity also depend on the planet's semi-major axis. The strongly misaligned cases have an abrupt eccentricity increase accompanied by an increase in eccentricity dispersion. We will explain why this is so in Section \ref{sec:planet_evol}.
\rev{For the cases with low initial mutual inclination, i.e. the cases with $i_{\rm pl}=0\degree$ and $30\degree$, the eccentricity evolution due to the planet should be minimal because the planet's forcing eccentricity ($e_{\rm forced} \propto e_{\rm pl}$) is practically zero. Consequently, the observed smooth growth in eccentricity and dispersion is by and large due to self-stirring processes in discs \citep{Ida+1993,Krivov+2018}.} 

Looking at Fig. \ref{fig:disc_res}, one can also see that the polar cases show the maximum value for the debris eccentricity $e_{\rm d}$ in every simulated case. The cases $i_{\rm pl}=\{30\degree,60\degree\}$, on the other hand, exhibit a significant difference between them, except in the case $a_{\rm pl} = 0.3$ au when both increase in eccentricity at practically the same rate as for $i_{\rm pl}=0\degree$. The case $i_{\rm pl}=30\degree$ shares the same behaviour as the case $i_{\rm pl}=0\degree$ in all the simulations. On the other hand, the case $i_{\rm pl}=60\degree$ shows an abrupt growth in the eccentricity in a short period accompanied by a large dispersion, similar to the polar case. \rev{These observations suggest a change in the dynamical behaviour below and above some critical angle between 30$\degree$ and 60$\degree$, as will be analyzed below. }

\begin{figure*}
\centering
\begin{center}
 \begin{tabular}{cc}
    \includegraphics[width=0.48\textwidth]{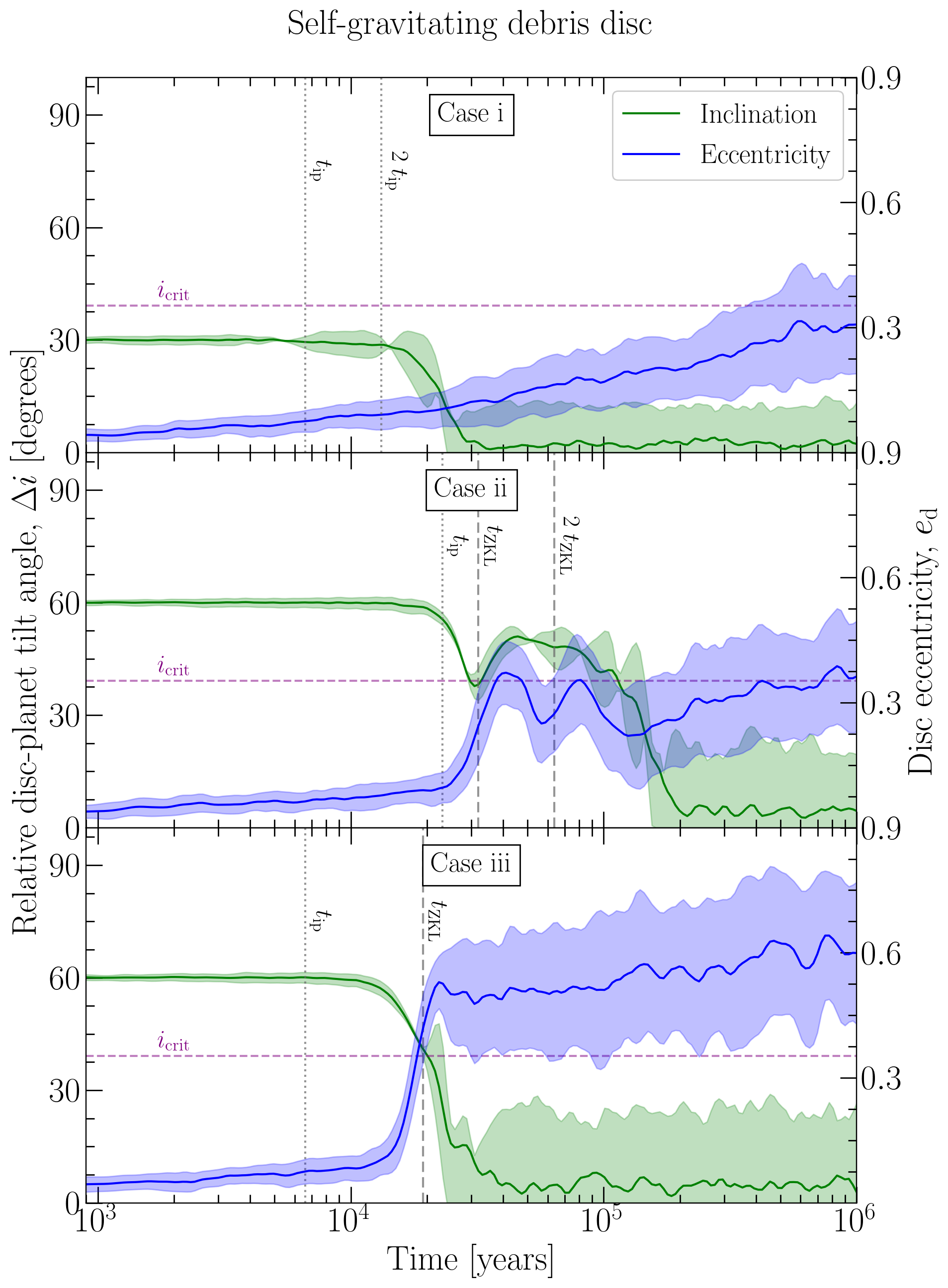} & \includegraphics[width=0.48\textwidth]{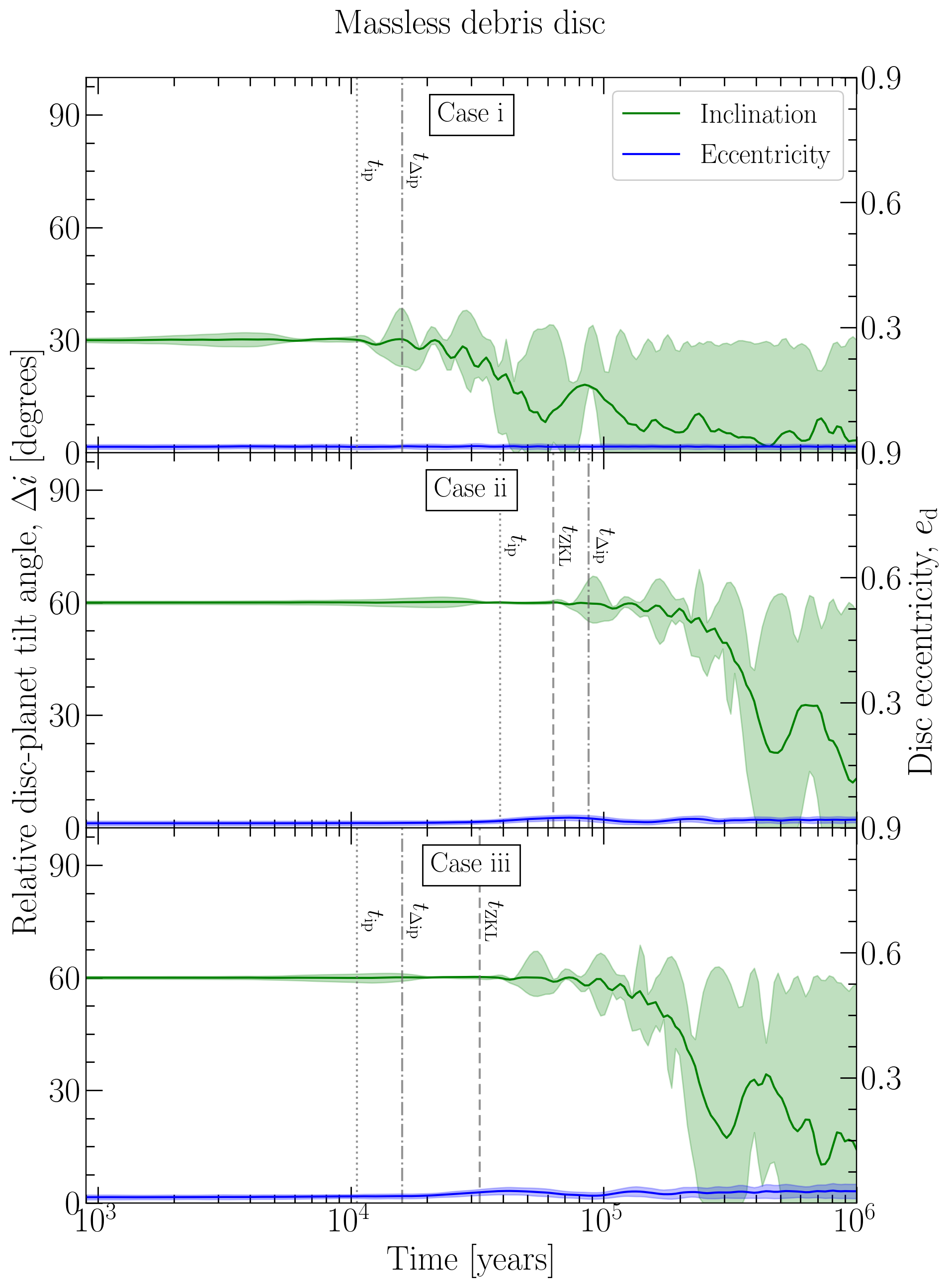} 
\end{tabular}
    \caption{Evolution and comparison between the characteristic debris eccentricity (in blue) and the relative tilt angle between the planet and the disc $\Delta i$ (in green) for the three different cases. Case~(i) represents the simulation with $a_{\rm pl}=0.7$ au and $i_{\rm pl} = 30\degree$, Case~(ii) is for $a_{\rm pl}=0.5$ au and $i_{\rm pl} = 60\degree$, and Case~(iii) is for $a_{\rm pl}=0.7$ au and $i_{\rm pl} = 60\degree$. \textit{Left-hand panels} show the results for a self-gravitating debris disc. The horizontal purple dashed line represents the value of $i_{\rm crit} = 39.2\degree$. The vertical grey dashed lines represent the ZKL timescale for each ZKL case ($t_{\rm ZKL}$), which is computed using Eq. \ref{appeq:t_KL}; this line is displayed twice in the middle panel representing one and two ZKL timescales. The grey dotted vertical lines represent the inclination precession timescale ($t\sbs{ip}$), which is computed using Eq. \ref{eq:prec-full}, and it is plotted twice in the Case~(i) panel. \textit{Right-hand panels} show the results for a massless disc. The grey dotted vertical lines also represent the inclination precession timescale, and the vertical dashed lines represent one ZKL timescale which is computed using Eq. \ref{appeq:t_KL_massless}. Additionally, the grey dash-dotted lines represent the differential inclination precession ($t_{\Delta \text{ip}}$) of the near and far disc edge, which is computed using Eq. \ref{eq:prec-diff}. }
    \label{fig:KL}
\end{center}
\end{figure*}

\subsubsection{Inclination-eccentricity correlations}
\label{sec:correlation}

Fig. \ref{fig:disc_res} shows that the abrupt changes of the debris' eccentricity and $\Delta i$ evident in some simulations occur practically simultaneously. Additionally, these abrupt changes are present in the simulation with $i_{\rm pl}=60\degree$ or $90\degree$, \rev{but not for $i_{\rm pl}=0\degree$ or $30\degree$}. Coupling between eccentricity and inclination for relatively large inclinations suggests the presence of von Zeipel--Kozai--Lidov oscillations (hereafter ZKL) \citep{Zeipel1910,Kozai1962,Lidov1962}. 

\rev{ZKL oscillations occur in hierarchical three-body systems when the  two constituent minor bodies have a relatively large mutual orbital inclination. This effect is characterized by a periodic exchange between inclination and eccentricity in one or both bodies. Depending on the three-body system architecture, we can distinguish between two configurations, mainly the classical and inverse ZKL. The classical ZKL refers to an outer body perturbing an inner body with a mass ratio of $m_{\rm out}/m_{\rm in}\gg1$, and the inverse ZKL refers to an inner body perturbing an outer body with $m_{\rm out}/m_{\rm in}\ll1$. Additionally, the classical ZKL appears only when the inclination of the outer body with respect to the inner body orbit exceeds a threshold of $i_{\rm crit} = 39.2\degree$ for prograde cases \citep{Kozai1962,Jefferys&Moser1966}, although other works considering single planets interior to discs suggest $i_{\rm crit} \sim 20\degree$ \citep{Terquem&Ajmia2010,Teyssandier2013}. On the other hand, the inverse ZKL requires a minimum mutual inclination of $i_{\rm crit} \sim 63\degree$ \citep{Vinson+2018,deElia+2019} if an inner perturber with a circular orbit is considered. A complete review of the ZKL and its applications can be found in \citet{Naoz2016} and \citet{Ito+2019}. We remark that, in our case, there is no such clear hierarchy because the planet and the disc are of comparable mass, and they can perturb each other significantly. Nevertheless, in this scenario, the planet would, by and large, evolve due to the classical ZKL, while the disc could undergo inverse ZKL oscillations. From hereon in, for simplicity, we will refer to both types as ZKL.} 

Results of Fig. \ref{fig:disc_res} suggest an eccentricity-inclination exchange like ZKL in our simulations, with eccentricity-inclination coupling present in models with $i_{\rm pl}\geq60\degree$ and absent from those with $i_{\rm pl}\leq 30\degree$. Nevertheless, there is no long-term periodic exchange between inclination and eccentricity as suggested for models with test particles \citep{Naoz+2017}. The inclination reaches a constant value at the end instead, likely due to the effect of self-gravity. \rev{Besides, our simulations suggest a critical angle closer to the classical ZKL value of $39.2\degree$ instead of $\sim20\degree$ or $\sim63\degree$. For this reason, we adopt the value $i_{\rm crit}=39.2\degree$ hereafter.}

In our simulations, we have shown that the disc tends towards being quasi-coplanar with the planet. However, the way in which they reach such a state differs between simulations. We identify three distinct scenarios from our simulations. The left-hand panels of Fig. \ref{fig:KL} show an example for each scenario. Case~(i) is the simulation with $a_{\rm pl}=0.7$ au and $i_{\rm pl} = 30\degree$, Case~(ii) is for $a_{\rm pl}=0.5$ au and $i_{\rm pl} = 60\degree$, and Case~(iii) is for $a_{\rm pl}=0.7$ au and $i_{\rm pl} = 60\degree$. Every case exhibits a different behaviour for the debris' eccentricity and $\Delta i$ as follows:

\begin{itemize}
    \item Case (i): in this case, the debris' eccentricity always increases gradually due to the process of self-stirring. Meanwhile, $\Delta i$ decreases (gradually or abruptly, depending on $a_{\rm pl}$) until it becomes practically zero. This case occurs for all simulations with $\Delta i \leq30\degree$. Note the angle is lower than $i_{\rm crit}$; therefore, no ZKL is expected in these cases. Exceptionally, in $\Delta i \leq60\degree$ for $a_{\rm pl}=0.3$ au, we observe a similar behaviour despite being in the ZKL regime. 
    
    \item Case (ii): in this case, the debris' eccentricity and $\Delta i$ experience a period without abrupt changes at the beginning. Then, $\Delta i$ decreases until it reaches a value close to $i_{\rm crit}$, before starting to increase. After a brief period, it decreases again, passing below $i_{\rm crit}$, and finally becomes practically zero. The eccentricity also manifests a period of oscillations, gradually increasing by the end. This stage can potentially be interpreted as damped ZKL oscillations. 
    To show this, one can use the \rev{planet-disc prescription along with a disc with homogeneous surface density approximation to compute the timescale (see Appendix \ref{app:KL}, Eq. \ref{appeq:t_KL}), these are shown using vertical black lines in Fig. \ref{fig:KL}. Doing so, we find that the timescales match with the onset of the inclination-eccentricity exchange. Additionally, it is noteworthy that the oscillations period in the $\Delta i$ dispersion are comparable with the estimated ZKL timescale as well as the eccentricity oscillations; see the middle-left panel of Fig. \ref{fig:KL}.} 
    
    \item Case (iii): in this case, the debris' eccentricity and $\Delta i$ experience a single abrupt exchange which happens at about half the ZKL time. The debris' eccentricity increases quickly, and $\Delta i$ decays to roughly zero (see the bottom-left panel of Fig. \ref{fig:KL}.). At the same time, the dispersion of both quantities increases. 
\end{itemize}

By the end, all cases that experience ZKL develop high dispersion in debris eccentricity $e_{\rm d}$ and relative inclination $\Delta i$. Note, however, that even though particles attain coplanarity with the disc from a median point of view, there is a large dispersion along the vertical plane. The implications of this for the disc morphology will be discussed in Section \ref{sec:morphology}. Nevertheless, it is important to note that the dispersion is considerably less than in the massless case, as we will describe in the next Section. 

\subsection{Self-gravitating vs. massless discs} \label{sec:self-gravity disc}

In order to highlight the impact of self-gravity in our simulations, we display in the right column of Fig. \ref{fig:KL} the results corresponding to the left column but now by switching off the disc gravity, i.e., using massless test particles.

Looking at Fig. \ref{fig:KL}, it is evident that the debris' eccentricity remains close to 0 in the massless simulations. This can be understood by the fact that the planetary orbit remains practically circular, and by construction, the disc does not self-stir itself. However, it is possible to observe a small growth in the debris' eccentricity in cases~(ii) and (iii). Additionally, the planet's inclination stirs the particles' inclinations strongly. The final inclination is distributed almost homogeneously in the range $\Delta i \leq i_{\rm pl}$, which corresponds to debris inclinations between 0 and $2 i_{\rm pl}$. This is expected for test particles, with the debris mid-plane tending to that of the planet \citep{Wyatt+1999,Pearce&Wyatt2014}. Therefore, by considering the self-gravity, it is evident that disc particles reduce their vertical dispersion by at least half compared to the massless case, i.e., $\Delta i < i_{\rm pl}/2$, and increase their eccentricities. In other words, both self-gravitating and non-self-gravitating discs will align their midplane with that of the planet, but self-gravitating discs will have a much smaller vertical extent around this plane than non-self-gravitating discs. \rev{Additionally, it is important to note that the final planet-disc plane is not the initial plane of the planet or the disc when self-gravity is accounted for (see Section \ref{sec:planet_evol}). }

We can examine the effect of self-gravity quantitatively by comparing the secular timescales. In a system with a narrow disc and a perturber on a near-circular and coplanar orbit, the timescales for secular evolution can be derived from Laplace--Lagrange theory \citep[see, e.\,g.,][]{Murray&Dermott1999}. The timescales for the evolution of orbital inclinations (inclination precession) are given by Eq.~\ref{eq:prec-full} in Appendix~\ref{app:np}.
For a massless disc with a central radius of $1.05$~au perturbed by a planet of $10^{-4}~M_\odot$ at $a_{\rm pl}=\{0.3, 0.5, 0.7\}$~au, the resulting periods are
\begin{equation}
   t\sbs{ip} (m\sbs{disc} = 0) = \left\{150, 39.0, 10.6\right\}\,\text{Kyr},
\end{equation}
respectively. These values are plotted as vertical dotted lines in the right-hand panels of Fig. \ref{fig:KL}. When the disc adds significant mass to the system, the timescales are shorter, explaining the quicker dynamical evolution observed for massive discs. Indeed, we find that
\begin{equation}
   t\sbs{ip} (m\sbs{disc} = 16.6\,M_\oplus) = \left\{77, 23,6.6\right\}\,\text{Kyr}.
\end{equation}
These values are plotted as vertical dotted lines in the left-hand panels of Fig. \ref{fig:KL}. 

It is worth mentioning that the oscillations in the dispersion in  Case~(i) show periods on the order of the inclination precession timescale. The corresponding timescales for differential inclination precession between the innermost and outermost disc edges are (Appendix~\ref{app:np}, Eq.~\ref{eq:prec-diff})
\begin{equation}
   t_{\Delta \text{ip}} (m\sbs{disc} = 0) = \left\{405, 87.0,16.0\right\}\,\text{Kyr},
\end{equation}
and they are plotted as vertical dash-dotted lines in the left-hand panels of Fig. \ref{fig:KL}.

The perturber can induce a complete misalignment of orbital planes across a massless disc on these differential timescales, i.e., $t_{\Delta \text{ip}}$. The timescales show a strong dependence on the separation between disc and perturber, with a factor of five between results for individual values of $a_\text{pl}$. That factor of five is also visible as a horizontal offset between the curves for $i_\text{pl} = 30^\circ$ in Fig.~\ref{fig:disc_res}. A direct graphical comparison with these analytical approximations is possible for the massless Case~(i) and Case~(ii) (Fig.~\ref{fig:KL}, right panels), where the computed 16~Kyr and 87~Kyr match the first maximum in inclination spread, respectively. For higher values of inclination, ZKL timescales become more important; for mass-bearing discs, timescales decrease, and interaction becomes more complex \citep[e.\,g.,][]{Pearce&Wyatt2014,Pearce&Wyatt2015,Sefilian+2021,Sefilian+2023}. However, it is intriguing that the match happens in the massless Case~ii but not in the massless Case~iii; this could be due to the combination of higher-order terms (eccentricity and inclination) in the secular timescales for high inclinations. 


\subsection{Planet evolution}
\label{sec:planet_evol}

\begin{figure}
\centering
\begin{center}
    \includegraphics[width=0.450\textwidth]{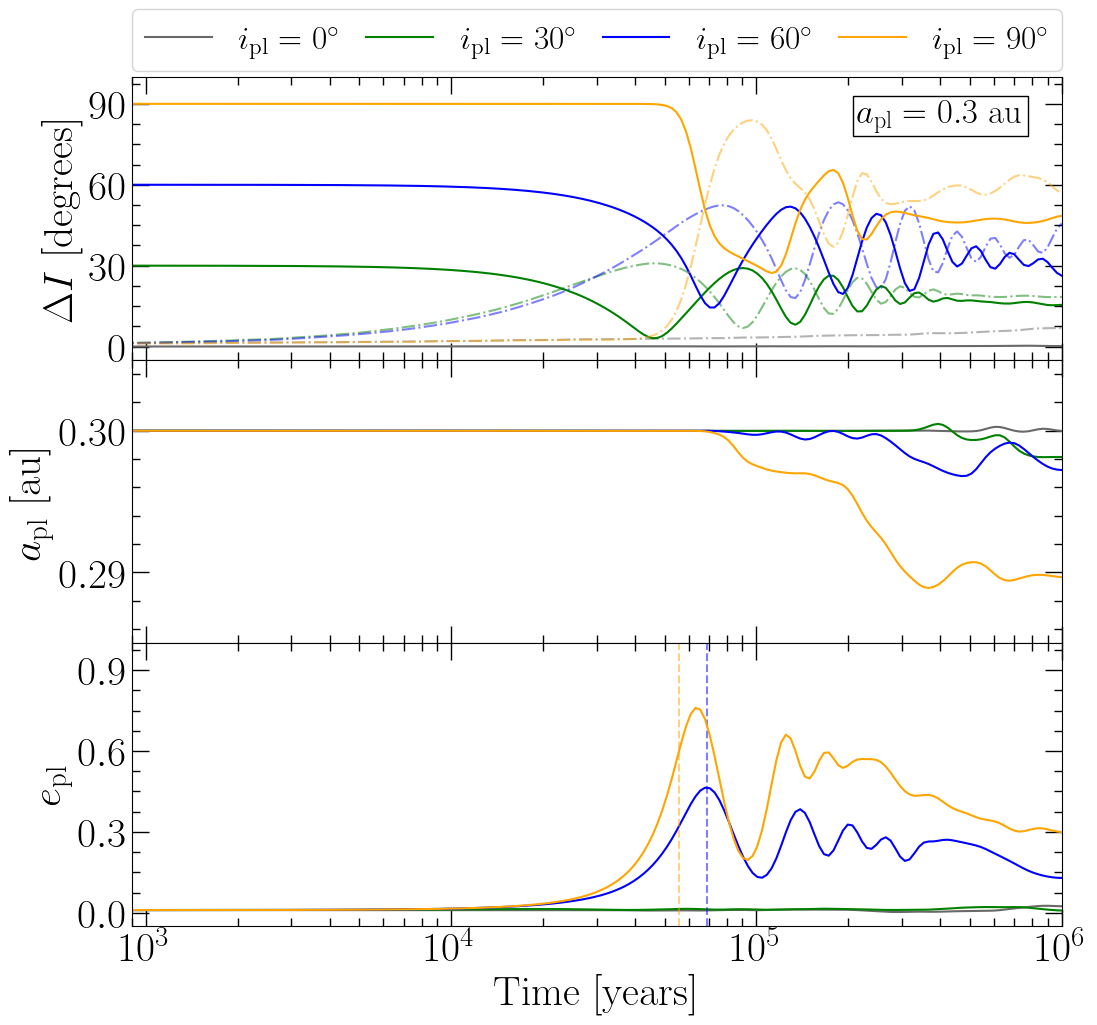}\\
    \includegraphics[width=0.450\textwidth]{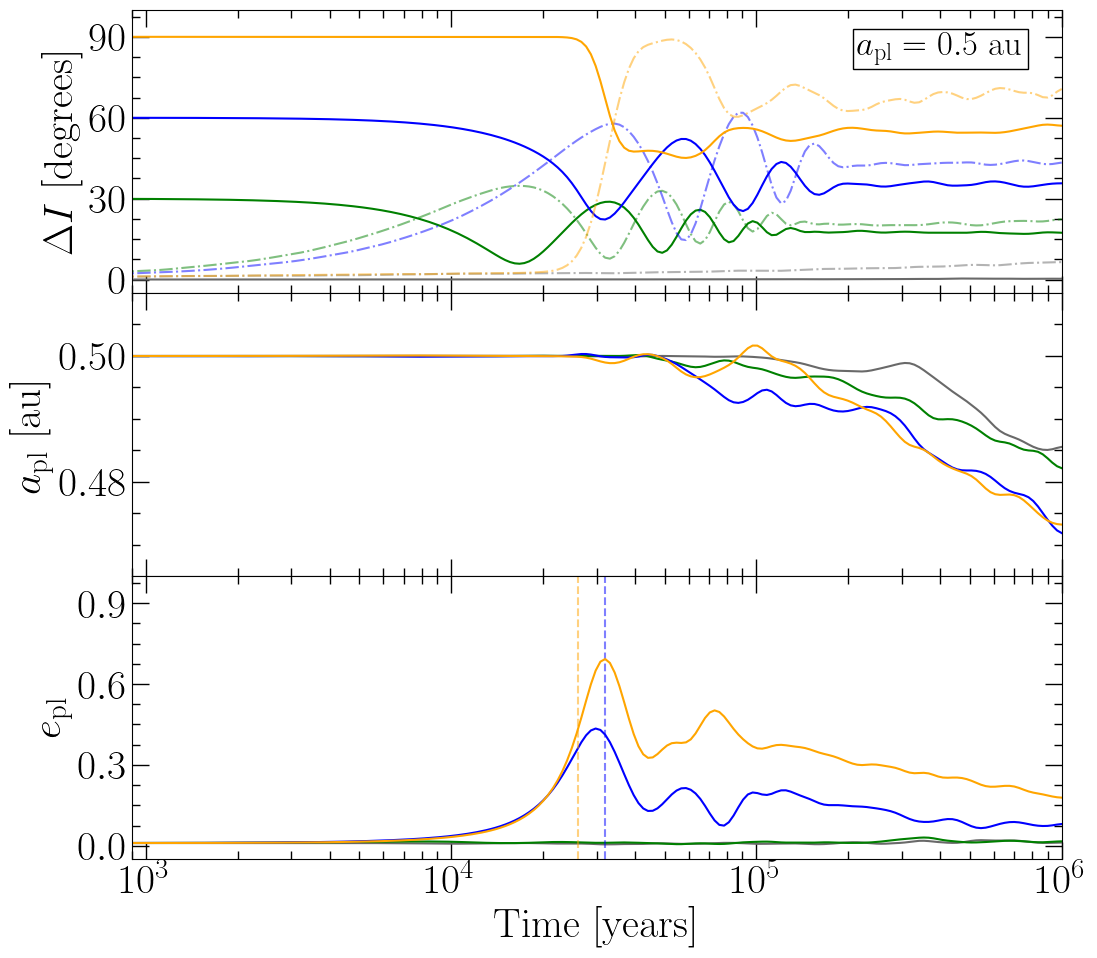}\\
    \includegraphics[width=0.450\textwidth]{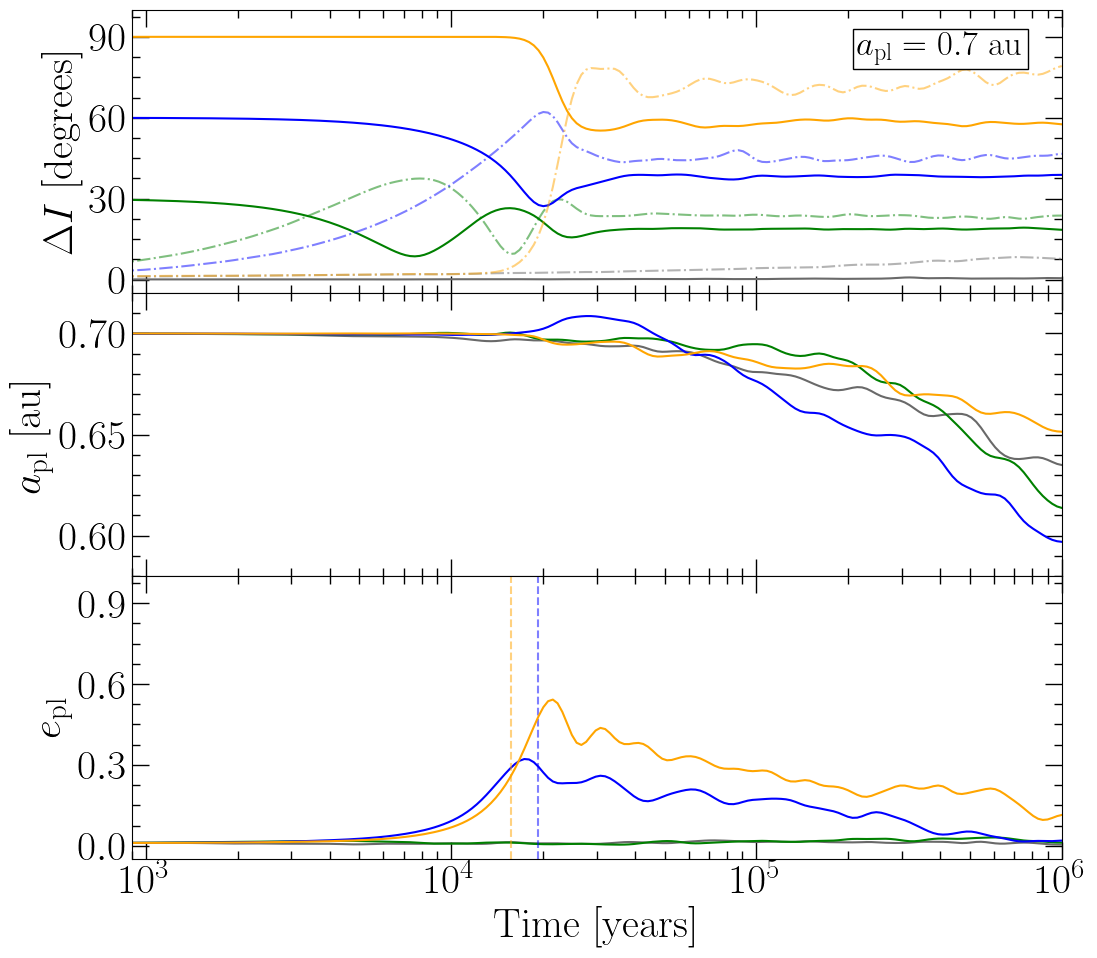} 

    \caption{Evolution of the planet inclination with respect to the disc's initial plane $\Delta I$ (top sub-panels), its semi-major axis $a_{\rm pl}$ (middle sub-panels), and its eccentricity $e_{\rm pl}$ (bottom sub-panels) for different initial semi-major axes of the planet. The colours represent the initial planet-disc orbit inclination ($i_{\rm pl}$). \rev{The dotted-dashed curve in every top sub-panel represents the disc inclination with respect to its initial plane. The vertical dashed lines in every bottom sub-panel represent $t_{\rm ZKL}$, which is computed using Eq. \ref{appeq:t_KL}.}}
    \label{fig:planet}
\end{center}
\end{figure}

We now focus on the evolution of the planetary orbit. Fig.~\ref{fig:planet} displays the evolution of the planet's inclination relative to the disc's initial midplane, $\Delta I$. As in the disc results, the planet's behaviour changes when the ZKL acts. 


Generally, the planet's inclination decreases over time in all simulations until a value close to 60\% of its initial value is reached; then stays there without significant changes. \footnote{Our choice to set the disc mass to half of the planet probably led to such a final value close to 50\%. Different mass ratio values would perhaps affect the final value, with a lower-mass disc expected to change the planet's inclination less.} In this process, the planet's inclination exhibits a damped oscillation even if the initial inclination value is below $i_{\rm crit}$. This said, though, we note that the polar case does not show significant oscillations compared to the other inclined cases. This damped behaviour \rev{correlates with the disc's inclination (with respect to its initial midplane) in an anti-correlated way until they converge in a quasi-coplanar configuration; see the solid and dotted-dashed lines in top-panels of Fig. \ref{fig:planet}}. 


For the semi-major axis evolution, we note a net inward migration in practically all cases, including the initially coplanar cases. This migration starts when the planet and/or disc particles acquire eccentricity. This leads to more scattering events between the planet and particles. When the simulation reaches 1 Myr, the remaining particles with a semi-major axis lower than the arbitrary threshold of 100 au are between 70\%-100\% for the non-ZKL cases and between 30\%-60\% for the ZKL cases with respect to the initial amount. \rev{Therefore, there is an energy exchange between the planet and particles led by the scattering processes aside from an angular moment exchange. Due to the ejected particles from the disc's local neighbourhood, the system's total energy and angular momentum are not conserved by the end.} While the planet can experience an inward or outward migration, the former dominates in our case because the planet is massive enough to eject planetesimals in close encounters. The planet migration driven by planetesimal scattering is studied in more detail in, e.\,g., \citet{Ida+2000}, \citet{Kirsh+2009}, and \citet{Friebe+2022}. In this process, the planet's orbit is also progressively circularized.


Finally, we can distinguish two cases in the evolution of the planet's eccentricity depending on the initial mutual inclination. For planets with initial inclinations larger than the critical inclination ($i_{\rm crit}<i_{\rm pl} = 60\degree, 90\degree$), their eccentricities show similar behaviour; first undergoing an abrupt increase, followed by a more gradual decrease (along with some oscillations). \rev{Note that the planet reaches its maximum eccentricity on a timescale comparable to the ZKL timescale. The small difference between when this happens and the ZKL timescale computed using Eq. \ref{appeq:t_KL} can be attributed to the fact that the latter assumes a solid disc, while our simulations include the ejection of particles from the disc. This can be noted in the case with $i_{\rm pl} = 60\degree$ and $a_{\rm pl} = 0.3$ au, where there is a practically perfect match between theoretical and simulated timescale values due to the ejection of particles from the disc being minimum.} The final planet eccentricity $e_{\rm pl}$ trend indicates that the large eccentricities that arise in the ZKL cases may eventually go to zero. Conversely, planets starting with an inclination lower than $i_{\rm crit}$ remain on low-eccentricity orbits, further supporting our explanation of the ZKL-induced eccentricity excitation.

A qualitative similar scenario was explored by \citet{Bitsch2013} but in the context of a planet embedded in a protoplanetary disc. Their results for the planet's eccentricity and inclination evolution are similar to ours. Additionally, \citet{Xiang-Gruess&Papaloizou2013} explored a similar scenario; whilst they could not resolve a full Kozai cycle, their results are similar to our initial evolutionary stage. 

\rev{While we currently do not have a complete understanding of the physical processes underlying the damping seen in Figs. \ref{fig:KL} and \ref{fig:planet}, we speculate that it is in part related to the so-called process of resonant friction investigated recently by \citet{Sefilian+2023}. Resonant friction is a special case of the well-known dynamical friction, but rather than resulting from scattering, it follows from the gravitational coupling between a planet and a self-gravitating disc \citep[see also][]{Tremaine1998,Ward+2003,Hahn2007}. As demonstrated in these works, such a coupling leads to the redistribution of the initial angular momentum deficit within the system, namely from the planet to the disc, and in the process, both the planetary eccentricity and inclination damp exponentially over time \citep[see section 5.2 of][]{Sefilian+2023}. We expect that for the cases with $i_{\rm pl} = 30\degree$, the planet-disc convergence toward a quasi-coplanar configuration could be explained by the process of resonant friction; nevertheless, when larger planetary inclinations are considered, the coupled effects of ZKL and resonant friction must be considered as well.}


\subsection{Disc structures}
\label{sec:morphology}

\begin{figure*}
\centering
\begin{center}
    \includegraphics[width=0.9\textwidth]{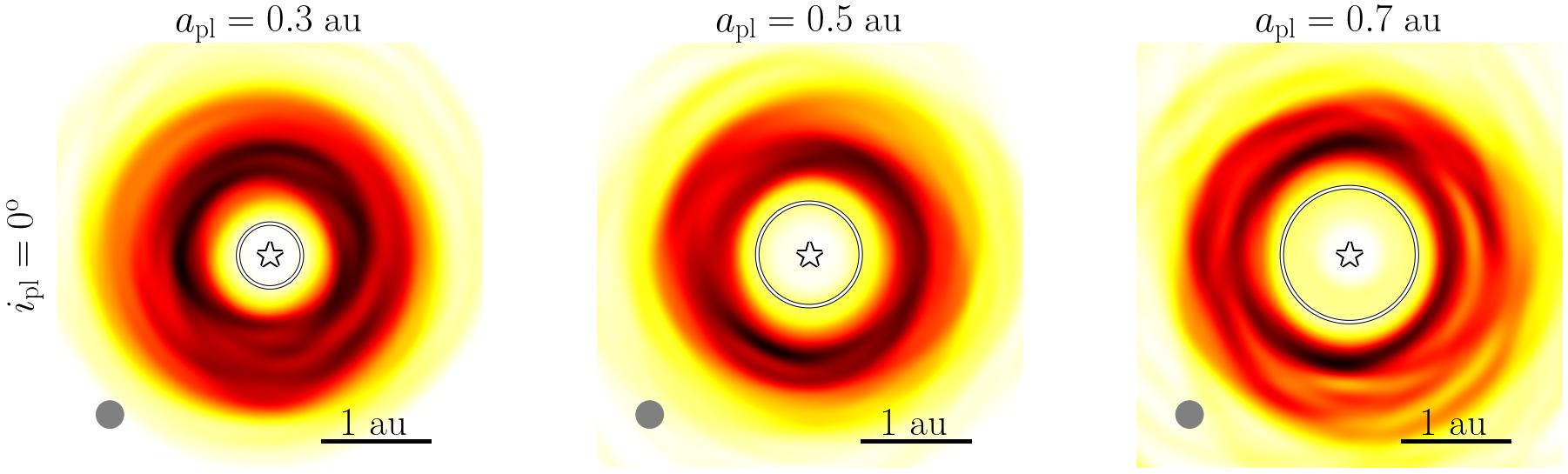}\\
    \includegraphics[width=0.9\textwidth]{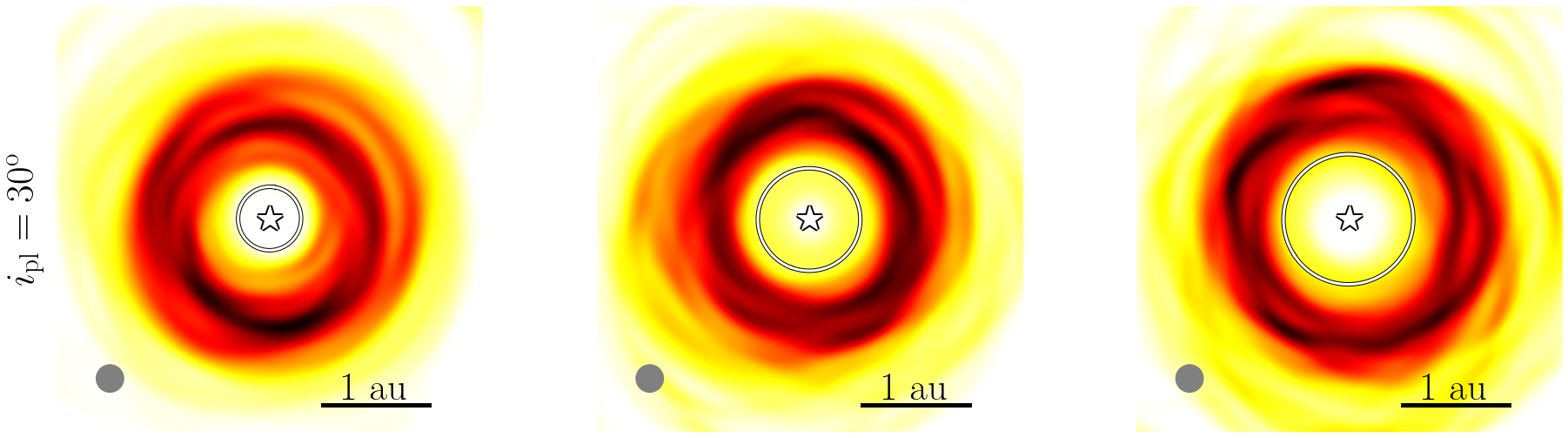}\\
    \includegraphics[width=0.9\textwidth]{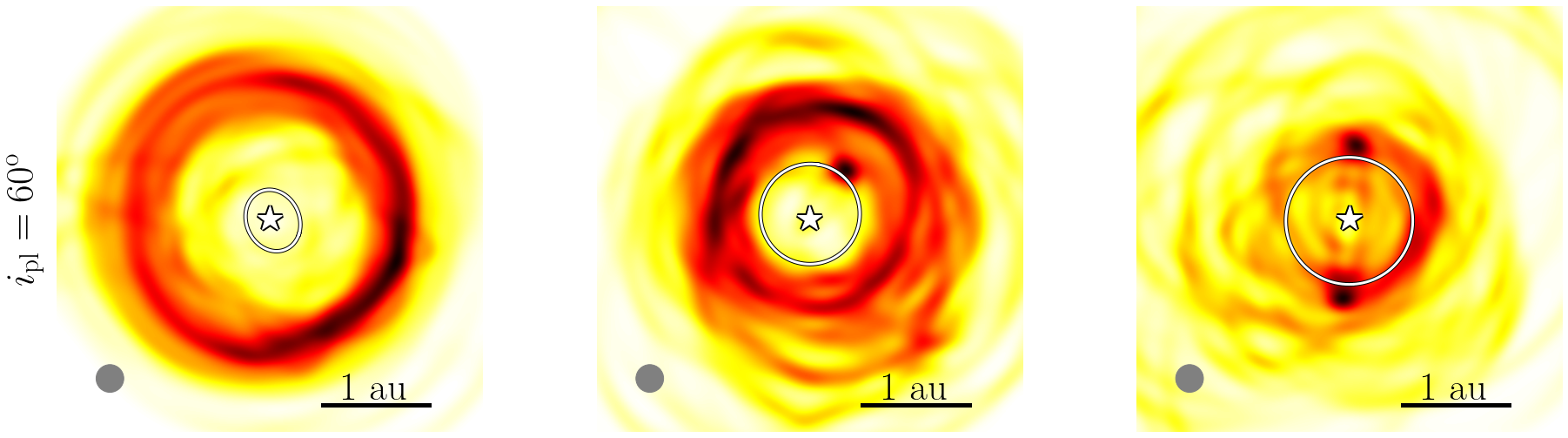}\\
    \includegraphics[width=0.9\textwidth]{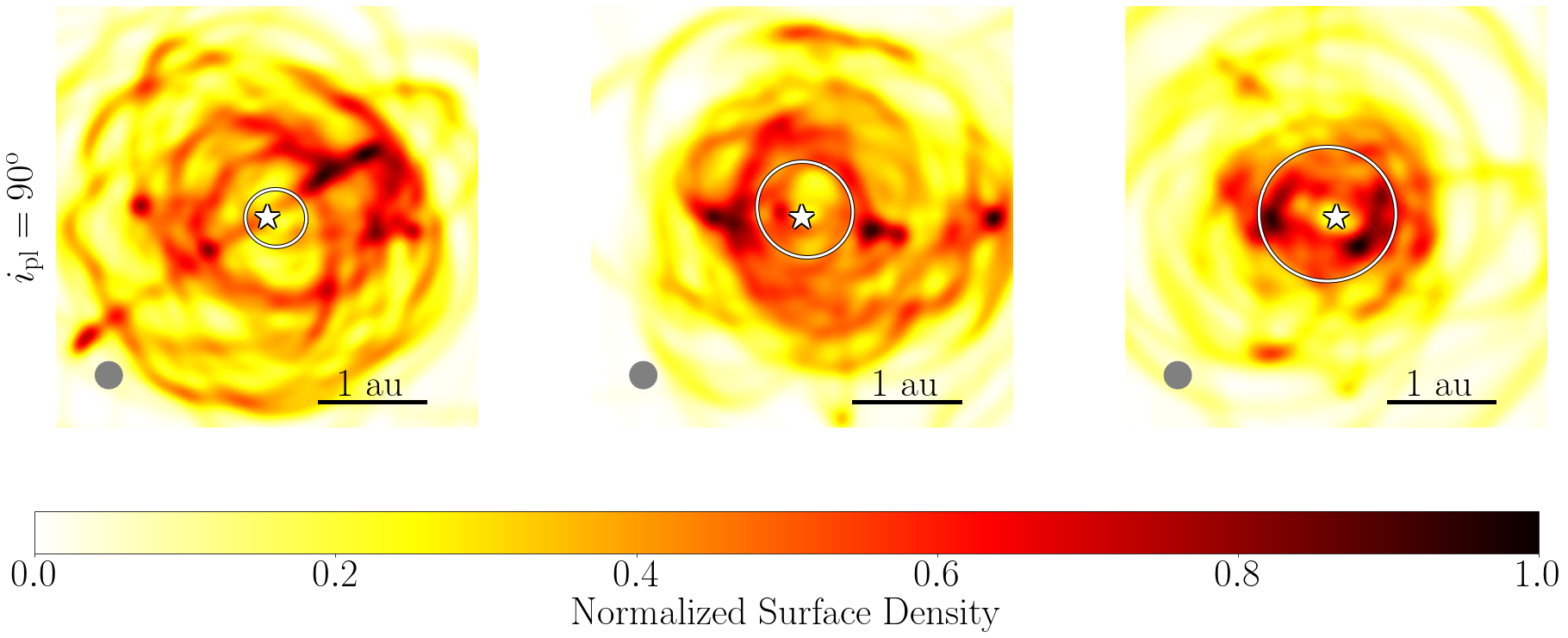}
    \caption{Face-on view of the debris disc's instantaneous surface density at 1 Myr. The disc is observed around the median angular moment vector of the disc (see Equation \ref{eq:disc_l}), which is orthogonal to the current disc plane. The different semi-major axes simulated are displayed in the three columns; meanwhile, the different inclinations are displayed in the rows. The solid white line denotes the planet's orbit. Each particle's orbit is populated with $10^3$ spawned particles with uniform distributed mean anomalies and then smoothed with a Gaussian filter of beam size that is displayed as a grey circle in every panel. \rev{In order to have a better resolution, we also averaged the last 50 snapshots before the final time simulated. These are averaged over the final 8 Kyr of the simulation.} Finally, the intensity has been normalized individually in each of the sub-panels. 
    }
    \label{fig:morphology_radial}
\end{center}
\end{figure*}

\begin{figure*}
\centering
\begin{center}
    \includegraphics[width=0.9\textwidth]{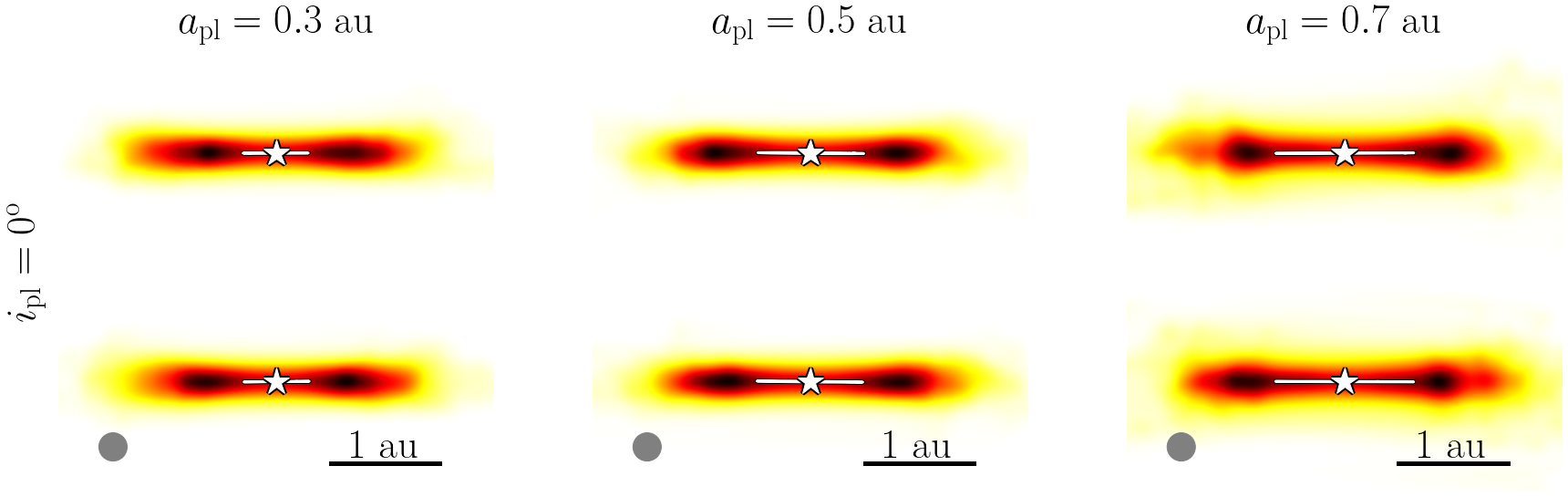}\\
    \includegraphics[width=0.9\textwidth]{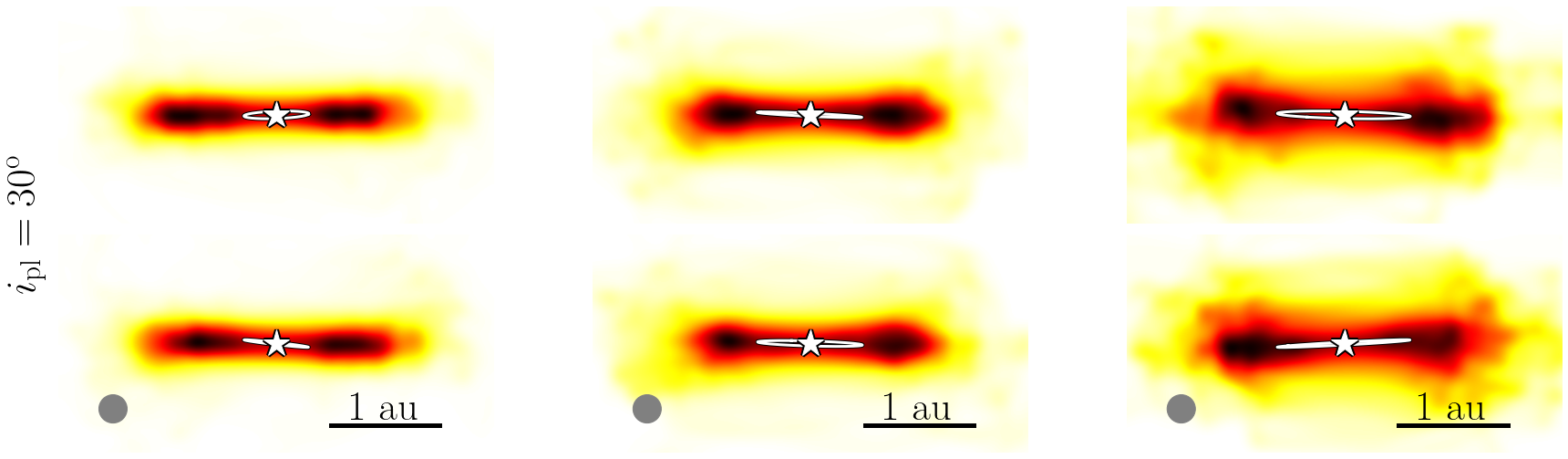}\\
    \includegraphics[width=0.9\textwidth]{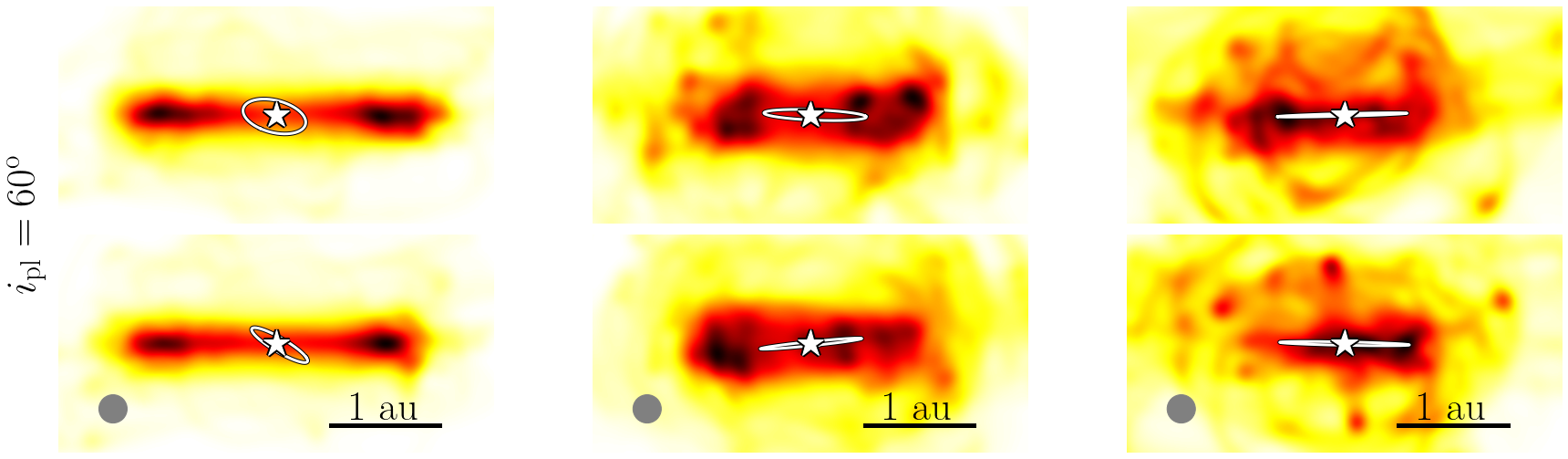}\\
    \includegraphics[width=0.9\textwidth]{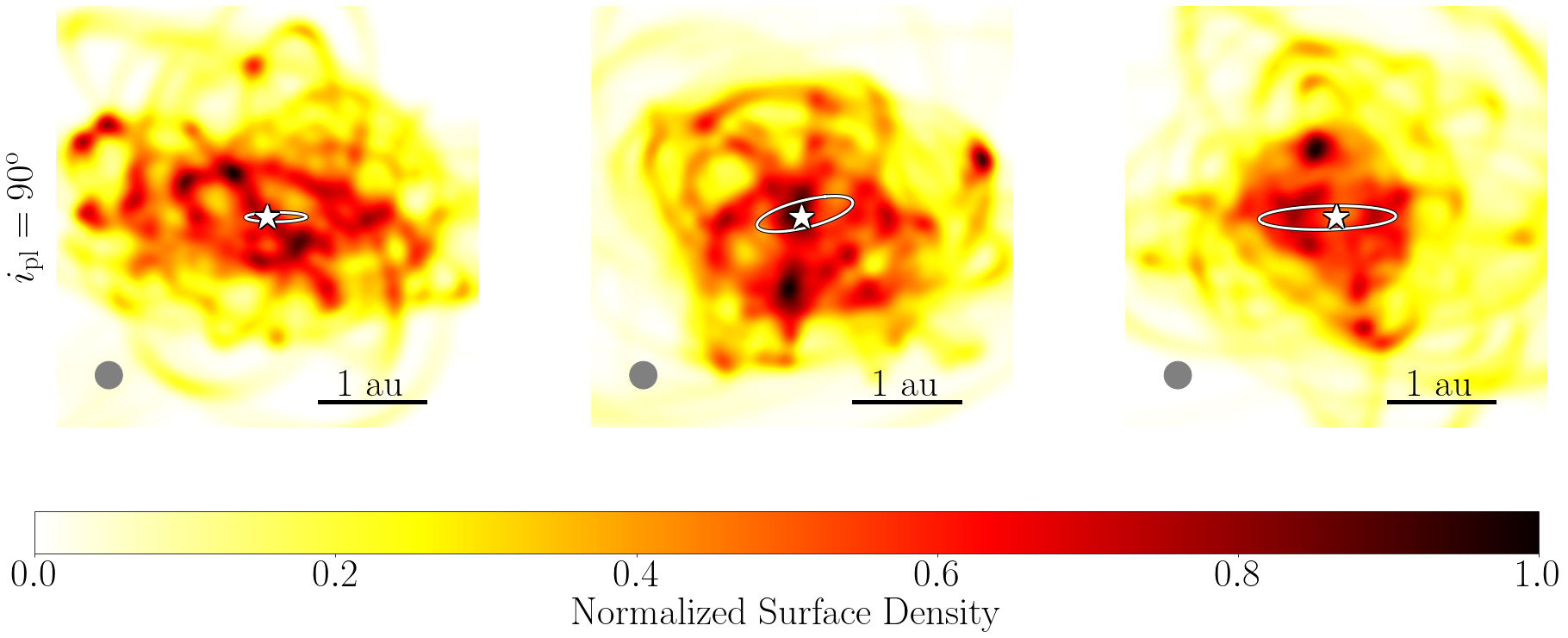}
    \caption{Same as Fig. \ref{fig:morphology_radial}, but for the edge-on view. The disc is viewed from its frontal and lateral sides for every panel, with the angular moment vector pointing upward to the current median disc plane. The polar case has only one view due to its level of disruption.}
    \label{fig:morphology_vertical}
\end{center}
\end{figure*}


We now explore the disc morphologies that result from the dynamical interaction studied thus far. Figures \ref{fig:morphology_radial} and \ref{fig:morphology_vertical} show the instantaneous surface density at 1 Myr for the radial and the vertical extensions, respectively. 

\subsubsection{Face-on view} 

The face-on view is shown in Fig. \ref{fig:morphology_radial}. On that view, we can see the radial extension of the disc. We can differentiate between two populations of dust at different radii. Most particles remain concentrated in a dense ring containing sub-structures. More strongly scattered particles populate regions more distant from that ring. The ring density is lower for larger initial values of $i_{\rm pl}$ because the scattering efficiency is enhanced when ZKL is present, leading to higher values of particles' eccentricity. However, the Case~(ii) ($a_{\rm pl}=0.5$ au and $i_{\rm pl} = 60\degree$) still shows a well-populated ring despite being in the ZKL regime, indicating that the presence of a ring also depends on the semi-major axis and not only on the planet's inclination.

As mentioned in previous sections, when ZKL is active, it produces high debris eccentricities, making particles populate the inner region more easily. Larger eccentricities trigger more close encounter events leading to an energy exchange and thus allowing some particles to migrate. In such cases, the particles reach distances equal to the planet's orbit or even a lower distance. The polar case, on the other hand, shows either a dense dust structure at the inner region of the planet's orbit or outside of it. The majority of the particles form a scattered cloud of particles. Particles that stay on near-circular orbits are just a minor fraction of the total. Most particles inside the planet's orbit are there due to their high eccentricities. 

In the ring, we can also observe arcs and clumps as sub-structures. Arcs are the most remarkable sub-structure when the ring holds many particles (see e.g., first, second and third row in Fig. \ref{fig:morphology_radial}). Meanwhile, clumps dominate as the main sub-structure when the ring becomes dispersed. Generally, arc features could be explained by pericentre glow \citep{Wyatt2005,Regaly2018}, while clumps arise due to the trapping of particles in mean-motion resonances with the planet \citep{Liou&Zook1999,Ozernoy+2000,Wyatt2003,Wyatt2006}. However, stochastic density variations due to the lower number of remaining particles in the high-inclination case are a more likely cause for the clumps.

\subsubsection{Edge-on view}

Fig.~\ref{fig:morphology_vertical} displays the disc structures in the edge-on view, unlike our massless simulations (see Fig. \ref{fig:KL}), our self-gravity simulations exhibit an important particle concentration in the mid-plane. The disc self-gravity is able to resist vertical shearing by the planet, so even a significantly inclined planet may not be able to generate strong vertical dispersion if the disc is sufficiently massive. Also, at the end of many of our self-gravitating simulations, the disc reaches a quasi-coplanar configuration with the planet, forming the ring described above. In analogy with the Kuiper belt, we can distinguish two different populations: cold and hot populations \citep[e.g.][]{Gladman&Volk2021}. The cold population comprises the particles that form the ring, and their inclinations remain in the quasi-coplanar state with the planet. Meanwhile, the hot population comprises scattered particles with high inclinations.

All cases with $i_{\rm pl} = 0\degree$, along with the cases with $i_{\rm pl} = 30\degree$ for {}$a_{\rm pl} = \{0.3,0.5\}$ au and the case with $i_{\rm pl} = 60\degree$ for $a_{\rm pl} = 0.3$ au look very similar. They preserve most of their particles in the mid-plane in a well-defined cold population. Additionally, they display two blobs at the disc's extremes, as expected in a ringed disc. The cases $i_{\rm pl} = 30\degree$ for $a_{\rm pl} = 0.7$ au, and $i_{\rm pl} = 60\degree$ for $a_{\rm pl} = 0.5$ au however, are more extended in the vertical direction with a larger hot population than the previous cases, as expected. Additionally, it is possible to observe inclined particles forming a warp sub-structure in the case $i_{\rm pl} = 30\degree$ for $a_{\rm pl} = 0.7$ au, which evokes the detected vertical structure of $\beta$~Pictoris \citep{Mouillet+1997,Heap+2000,Golimowski+2006}. Even though this aspect is subtle, it is the only remnant of the initial misalignment between the disc and the planet.

In the case with $i_{\rm pl} = 60\degree$ for $a_{\rm pl} = 0.7$ au, the inclined history of the planet-disc system is more evident. In this case, the most noticeable aspect is the depopulation of the cold population in favour of the hot population.
Although some particles form a ring in alignment with the planet's orbit, most particles contribute to the spreading of the vertical extension with an inclination larger than $i_{\rm crit}$. The extreme case is the polar case, which presents the most chaotic distribution, where particles were practically randomized in a sphere with only a few particles forming a ring, probably in resonant orbits. If, at some point, the particles reach inclination values lower than the critical value, the result will be particles with high eccentricities. The formation of a resonant ring after experiencing a disruption event from a more massive disc is studied in \citet{Pearce+2021} in the context of the Fomalhaut system.

\section{Discussion}
\label{sec:discussion}

So far, self-gravitating debris simulations using direct N-body codes have not been widely undertaken due to their computational cost, with a recent exception in \citet{Das+2023}. Previous works with a single inclined perturber interior to a massless disc have difficulty reproducing the thin debris discs observed, often leading to the conclusion that any unseen planets must be on coplanar orbits. However, our work shows that disc self-gravity could counter vertical shearing by an inclined planet, resulting in thin discs even if an initially inclined perturber is present (e.g. Fig. \ref{fig:KL}). The main consequence is the reduction of the spread of inclinations, as was shown in Fig. \ref{fig:KL}. We discuss the implications of this finding in more detail in Section \ref{sec:thickness}, before considering how our results may be extrapolated to radially wider discs (Section \ref{sec:wide_discs}).

\subsection{On the disc's thickness}\label{sec:thickness}

\begin{figure}
\centering
\begin{center}
    \includegraphics[width=0.48\textwidth]{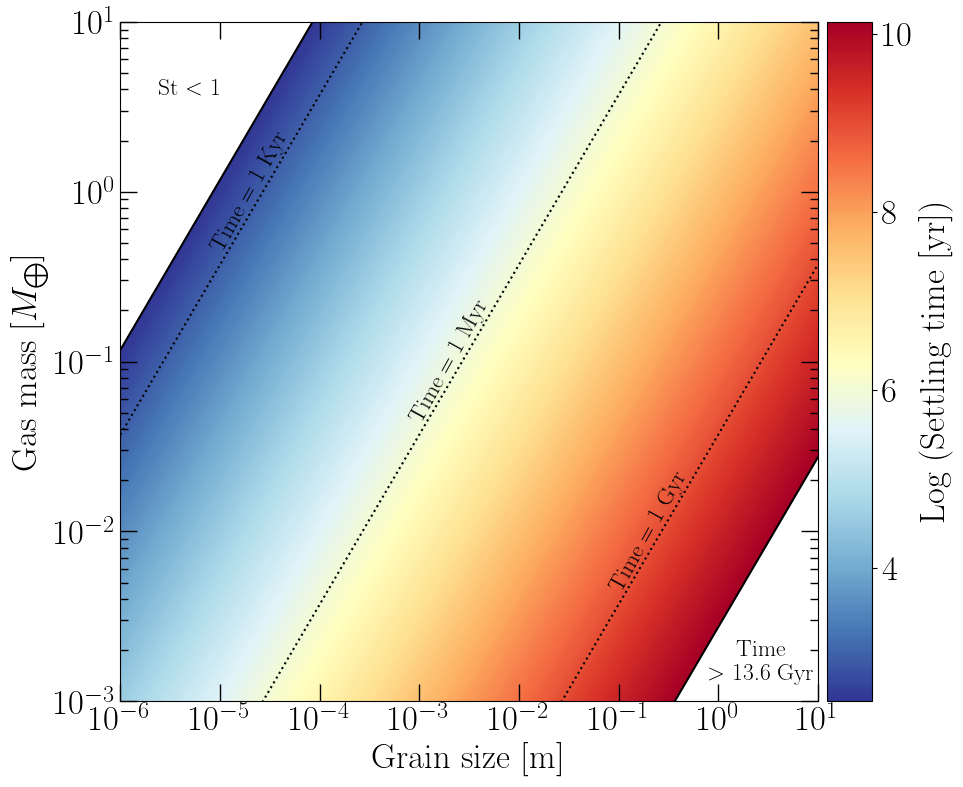} 
    \caption{\rev{Settling time as a function of the total gas mass and the grain size \rev{accordingly to Eq. \ref{eq:tau}}. The settling time is computed at the inner edge of a disc extending from 100 au to 110 au and assuming a constant gas surface density with a dust grain density of 3~g/cm$^3$. The white spaces in the upper-left and bottom-right corners represent the regime where the Stokes number is lower than 0 (our calculations would not be valid) and the times larger than the universe's age, respectively. The dotted lines represent isochronous.}}
    \label{fig:settling_time}
\end{center}
\end{figure}


Observations suggest that exoplanets populate an extensive range of orbital inclinations, from prograde to retrograde orientations \citep{Winn&Fabrycky2015}. Besides, in systems known to harbour debris discs, star-disc misalignment is commonly observed \citep{Watson+2011,Greaves+2014}, and more massive stars are prone to harbour planets with high inclinations \citep{Louden+2021}. These samples reflect planetary populations close to the star principally since planets in the outer regions are hard to observe and characterize. However, the current samples show the diversity in planetary systems. This observed diversity raises an important question on the presence of planets and discs found in inclined configurations. If so, this would imply that the systems with ZKL configurations should also be observed, with their associated sizeable vertical dispersion; \rev{based on our simulations, we got a minimum aspect ratio $h$ of $\sim0.1$ for initially coplanar configurations, while we obtained $h\sim0.6$ for the case with $i_{\rm pl} = 60\degree$ and $a_{\rm pl} = 0.7$ au, a case affected by ZKL. However, the high aspect ratios we predict are not borne out in observed debris discs; aspect ratios have been measured in several debris discs, with AU~Mic being the thinnest ($h=0.02$) and HD~110058 the thickest ($h=0.21$). Besides, the current sample of debris discs shows that about 66 \% of discs have an aspect ratio less than 0.1 \citep{Terril+2023}.}  

We can try three scenarios to approach the scarcity of vertically thick discs. \rev{The first possibility is that vertically thick discs are harder to image because the spreading of material in the vertical direction would mean that a puffed-up disc seen edge-on would have a lower surface brightness than a thin disc. For discs that are viewed face-on, the distribution of vertical thicknesses is poorly constrained. In addition, SEDs are affected only indirectly by the aspect ratio through differences in the size distribution caused by different impact velocities among dust grains \citep{Thebault&Augereau2007}.} Nevertheless, our simulations show that a disc can survive with a significant cold population even if a larger hot population exists. Only our polar simulations support this scenario since there is no coherent disc structure by the end.

The second scenario could be collisional depletion, which we neglected in this work. High orbital inclinations and eccentricities increase the relative planetesimal velocities and hence, impact energies and the fraction of disruptive collision. At the same time, collision rates are proportional to relative velocities. However, the discs' drastically increased horizontal and vertical extents reduce the particle number density and the collision rate with respect to the initial narrow belt. \rev{\citet{Lohne+2008} show collision timescales $\tau_{\rm col}$ to depend on eccentricities $e$ and radial extent ${\rm d} r$ as $\tau_{\rm col} \propto e^{2(1 - q_{\rm p})} {\rm d} r$, where ${\rm d} r \propto e$ for a puffed-up disc. For a primordial size distribution index $q_{\rm p} = 5/3 \ldots 11/6$ this results in $\tau_{\rm col} \propto e^{-1/3 \ldots -2/3}$. As a result, weak differences are expected for the timescales of the collisional evolution and the resulting disc depletion rates for our set of simulations; a weak impact on the abundance of thin versus thick discs is expected, with puffed-up discs becoming more and more depleted towards older, more evolved samples.}

The third scenario would be a mechanism that settles the particles in the mid-plane \rev{or keeps particles in the mid-plane when disruptive events (like ZKL) occur.} That may happen by the action of gas (which we discuss in Section \ref{sec:gas_drag}), the action of disc self-gravity (such as this work), or a combination of both. Under such a damping mechanism, debris discs in systems with highly inclined planets could end up looking like the thin discs in coplanar systems.

\subsubsection{Gas drag as an extra thinning mechanism}\label{sec:gas_drag}

\citet{Nakagawa+1986} computed the equations of motion for a dust particle embedded in gas. The vertical motion for a particle in a non-self-gravitating gaseous disc is given by:
\begin{equation}
\ddot{z} = - \left(\Omega_k{\rm St^{-1}}\right)\dot{z} -\Omega_k^2 z,
\end{equation}
where $\Omega_k$ is the Keplerian frequency, and St is the Stokes number. This linear ordinary differential equation has three solutions depending on the Stokes number. We use the regime where $\rm St\gg 1$, in which case the solution corresponds to a damped harmonic oscillator with an exponential decay time of $\tau = 2 \mathrm{St} \Omega_k^{-1}$: the settling time. See Appendix \ref{app:tau and St} for the whole derivation. The settling time can be expressed according to disc parameters as follows:
\begin{multline}\label{eq:tau}
    \tau = 5.885\ \left[\mathrm{yr}\right]\times \\  \left( \frac{\rho_s}{\rm g/cm^3} \right) \left( \frac{s}{\rm m} \right) \left( \frac{\Delta R}{\rm au^2} \right) \left( \frac{M_*}{ M_\odot}  \right)^{-\frac{1}{2}} \left( \frac{M_g}{ M_\oplus}\right)^{-1} \left( \frac{r}{\rm au}\right)^{\frac{3}{2}},
\end{multline}
where $\rho_s$ is the intrinsic dust grain density, $s$ is the grain size, $\Delta R = r_{\rm out}^2 - r_{\rm in}^2$: the difference between the squares of the outer rim and the inner rim of the disc, $M_g$ is the total mass of gas, $M_*$ is the mass of the star,  and $r$ is the radial distance. Fig. \ref{fig:settling_time} illustrates some values for the settling time as a function of the gas mass and the grain size for a narrow debris disc. The settling time estimation for the millimetre size is roughly ten million years or less, depending on the gas mass. The transition between a protoplanetary disc to a debris disc is subject to debate. Estimated ages of protoplanetary discs range from 2-3 Myr \citep{Espaillat+2017} to 40 Myr \citep{Zuckerman&Song2012}. Therefore millimetre grains should have enough time to settle. For late stages, i.e., in the debris phase, it would be necessary to consider second-generation gas. The gas is released via collision among grains, and misaligned configurations could lead to growing collision probabilities. Consequently, positive feedback could happen between the inclined orbit configurations and the amount of gas released. Nevertheless, exploring such effects is beyond the scope of this work. 


Recently, \citet{Olofsson+2022} explored the impact of gas on the vertical disc spread. They found that dust particles can efficiently settle toward the mid-plane if gas is present. This could explain the issue between the observation and our vertically spread simulations at their end state. Our simulations suggest a quasi-coplanar configuration between the angular moment of the planet and the disc by the end. By considering gas, it could be possible to settle the dust in order to reduce the inclination dispersion and make the disc thinner. 

Nevertheless, assuming the presence of gas in debris disc carries the question about its origin and the amount. If the gas came from the protoplanetary disc, so-called first-generation gas, then it would initially be abundant but short-lived \citep{Alexander+2006}. Alternatively, it can be released from grains in the late stages, the second generation. By considering  first-generation gas, we would assume that the event in which the planet acquired its high inclination happened close to the transition to a debris disc: from a gas-rich to a gas-poor disc. In the protoplanetary disc stage, it is possible to form planets with large eccentricities and/or inclinations \citep{Goldreich&Tremaine1980,Thommes+2003,Lega+2013}.
Consequently, there would be enough gas to make the disc thinner in that condition. On the other hand, if the planet became eccentric and/or inclined afterwards, there would be required second-generation gas to make the disc thin. In that scenario, the amount of gas would be considerably less, and hence the settling time will increase.

\rev{Table \ref{tab:taus} compares the gas content estimated via CO emission in HD~21997 and $\beta$~Pictoris, which harbour gas thought to be first and second generation, respectively. By estimating the settling time for millimetric dust, the second-generation gas would not be able to settle dust grain efficiently ($\tau_{(1\ {\rm mm})}\gg40$ Myr), but a gas content like the one measured in the HD~21997 system could. }

\begin{table}
\centering
\begin{tabular}{cccccc}
\hline
\hline
Name & Star mass & CO mass & $R_{\rm in}$ & $R_{\rm out}$ & $\tau_{(1\ {\rm mm})}$ \\
& [$M_{\odot}$]  & [$M_{\oplus}$]  & [au] & [au] & [yr]\\
\hline
HD~21997 & 1.80 & $5.0\cdot10^{-2}$ & 26 & 140 & $6.60\cdot10^5$ \\
$\beta$~Pictoris & 1.75 & $3.4\cdot10^{-5}$ & 20 & 120 & $4.91\cdot10^8$ \\
\hline
\end{tabular}
\caption{\rev{Estimation of the settling time due to gas drag for millimetric dust grain size ($\tau_{(1\ {\rm mm})}$) for two debris disc systems: HD~21997 and $\beta$~Pictoris. The disc parameters were taken from \citet{Kospar+2013} and \citet{Matra+2017}, respectively. We assume that all gas mass is the total CO mass, which is a reasonable approximation if the CO mass corresponds to the lower limit for the real total gas mass. We repeated the same exercise to make Fig. \ref{fig:settling_time}; we assume a constant disc surface density with extension from $R_{\rm in}$ to $R_{\rm out}$. }}
\label{tab:taus}
\end{table}



In summary, the observed population of relatively thin debris discs can be reached if we consider some mechanism, such as disc self-gravity (see Section \ref{sec:disc_evol}), gas drag (see Section \ref{sec:gas_drag}) or the joint action of both. As we showed in Fig. \ref{fig:settling_time}, gas affects the vertical grain settling differently, and different layers could appear.  Evidence of the vertical stratification of dust grains can be found in the AU Mic system \citep{Vizgan+2022}. Nevertheless, it would be necessary to have more information about the planets' misalignment to the star's spin, especially in distant orbits, to constrain this and future models, along with the impact of gas on debris disc structures. \rev{Additionally, it is worth mentioning the process of dust generation also plays an essential role in the vertical structure where the vertical dust distribution is also a function of the grain size \citep{Pan&Schilchting2012}. Hence the issue of the diversity of exoplanet inclinations compared to the thinness of observed debris discs remains a challenge to resolve.}


\subsection{Extrapolation to wide debris discs}\label{sec:wide_discs}

The initial setup of our simulations considers an initially narrow disc. It has a fractional width of $\Delta r/r \approx 0.1$, similar to the results obtained for millimetre-wavelength observation of Fomalhaut \citep{Kennedy2020} and HR~4796 \citep{Kennedy+2018} among other examples. Nevertheless, it is possible to find extended debris discs in the literature. One outstanding example would be HD~206893 \citep{Marino+2020,Nederlander+2021} with a fractional width larger than 1.0. Whether debris discs are born narrow or broad is an active area of research, and more studies are needed to unravel it. Thus it is reasonable to consider, as an initial condition, a broad disc instead of a narrow disc.

In a more extended radial configuration, the effect of the planet on the disc will depend more evidently on the distance. Inner regions will be affected faster than the outer regions of the disc. Therefore, the disc would be stratified radially according to the different time scales in which the planet affects them. When considering an eccentric planet, the effect can be appreciated graphically in coplanar configurations. In those cases, some rings and gaps could appear \citep{Pearce&Wyatt2015,Sefilian+2021,Sefilian+2023}.
Extrapolating this effect, but now for inclined configurations, could produce warped and/or tilted discs \citep{Mouillet+1997,Dawson+2011,Batygin2018}.

Although it is possible to observe warps in some debris discs, for example, $\beta$~Pictoris \citep{Mouillet+1997}, tilted and warped discs have been more extensively studied in the context of accretion discs in stellar and black hole environments \citep{Nelson&Ppaloizou1999, Lodato&Price2010, Facchini+2013}. Depending on the discs' viscosity, the warps' propagation can change, and even break the disc at some radius. In Section \ref{sec:results}, we commented on the similarities between our results and some works on protoplanetary discs. Therefore it is plausible to make an analogy between the two regimes. If inclined companions can trigger warps and tilted gaseous discs, something similar could happen in the debris discs, with a different timescale. By following the models, we could have an inner part in a quasi-coplanar orientation with the planet (similar to our simulations) and the outer part of the disc misaligned with respect to the inner disc.

Another aspect relevant to the extrapolation is the disc's mass. \rev{Although exploring the effect of the disc mass on the final state of the simulations is beyond the scope of this work, we can infer some possible scenarios.} We used an arbitrary value for the disc mass, although consistent with estimates of some extrasolar debris-disc masses (see Section \ref{sec:methods}), and the disc shape was conserved in many simulations. \rev{However, the planet exhibited changes in its orbit plane. With a larger disc mass value, the final planetary orbital plane should be closer to the initial disc plane, which would remain largely unchanged, than to the planet's original orbital plane.} In addition, the disc stirring could increase, affecting the disc structures. 



\section{Conclusion}
\label{sec:conclusion}

We performed N-body simulations of inclined and narrow debris discs with an interior planet, considering self-gravitating particles with a disc-to-planet mass ratio of 0.5. Our results reveal strong changes in comparison with the massless regime and that the final state for several configurations can be the same despite their initial conditions. Our main findings are
summarized as follows:

\begin{enumerate}[(i)]
    \item For most simulations, the final alignment between the planet and the disc mid-plane is quasi-coplanar (less than 10$\degree$) with relatively moderate associated dispersion around the final value. Setups with planets on initially polar orbits are the only ones that showed a complete disc disruption. \rev{Including self-gravity in simulations, the inclination dispersion was reduced compared to the massless simulations.} Additionally, in all cases, the debris eccentricity grows. \rev{This is attributed to the self-stirring processes in non-ZKL cases and a combined effect of self-stirring and ZKL when the latter acts.}  
    
    \item The Zeipel--Kozai--Lidov oscillations play an important role in the evolution of the planet-disc system if the planet is initially sufficiently inclined. The cases in this regime experience the most drastic increase in eccentricity and inclination dispersion. \rev{We can distinguish two evolutionary paths in the ZKL regime: (a) when the exchange between inclination and eccentricity happens in one step, and (b) when the exchange presents two or more oscillations before becoming stable by the end due to damping.}
    
    \item The planet reduces its initial inclination relative to the disc and migrates inwards in all cases. \rev{The configurations with the larger semi-major axis are affected more strongly by migration.} In addition, the planet acquires a large eccentricity when the Kozai-Lidov oscillations act. \rev{The planet reaches a maximum eccentricity value in a time comparable with a ZKL timescale, being circularized progressively afterwards.}
    
    \item Morphologically, the final states of the simulations look similar despite the different initial conditions, except when the planetary orbit is initially polar. The radial and vertical extensions depend on the planet's inclination and semi-major axis. \rev{The ZKL effect strongly changes the disc morphology. When ZKL acts, the radial and vertical extensions become wider. In the radial extension, the distribution of particles is broad due to their high eccentricity. Meanwhile, in the vertical direction, we find a particle population that is highly inclined and another that remains in the midplane.}
\end{enumerate}

Our work represents a step forward towards better understanding planet--debris disc interactions.  Our results show that a self-gravitating disc can survive undisrupted in the presence of an initially highly inclined planet, converging to an almost coplanar state at the end. This suggests caution in interpreting the dynamic history of the debris disc by only observing it since many configurations can end up looking alike. More detailed information about the whole system is needed to unravel its past; determining the misalignment of a planet (or planets) relative to the star's spin, or characterising vertical and radial profiles will become an essential aspect of doing that.

\section*{Acknowledgements}
P.P. Poblete, T. L{\"o}hne, and T.D. Pearce are supported by \textit{Deutsche Forschungsgemeinschaft} grants Lo 1715/2-2, Kr 2164/14-2, and Kr 2164/15-2. A.A. Sefilian acknowledges support by the Alexander von Humboldt Foundation through a Humboldt Research Fellowship for postdoctoral researchers.

\section*{Data availability}
The data underlying this article will be shared on reasonable request to the corresponding author. The code {\sc Rebound} used in this work is publicly available at \hyperlink{https://rebound.readthedocs.io/en/latest/}{{\sc Rebound} home page}.

\bibliographystyle{mnras}
\bibliography{paper}

\appendix

\section{Zeipel--Kozai--Lidov timescales}
\label{app:KL}
\rev{The Zeipel--Kozai--Lidov timescale is computed for two different scenarios: (i) for an outer massless particle perturbed by an inner planet and (ii) for a planet perturbed by an outer massive disc.} 

\begin{itemize}

\item[(i)] \rev{ Massless particles: the timescale is taken from equation 22 of \citet{Naoz+2017}. It gives the timescale for an outer test particle as follows:}
\begin{equation}\label{appeq:t_KL_massless}
    t_{\rm ZKL} = \frac{4}{3}T_{\rm pl}\left(1-e_{\rm pl}^2\right)^2\frac{\left(M_*+m_{\rm pl}\right)^2}{M_* m_{\rm pl}}\left(\frac{a_{\rm disc}}{a_{\rm pl}}\right)^2,
\end{equation}
\rev{where $M_*$ and $m_{\rm pl}$ are the mass of the central star and the planet, respectively. $T_{\rm pl}$ is the period of the planet, and $a_{\rm pl}$ is the semi-major axis of the planet, while $a_{\rm disc}$ is the semi-major axis at the middle of the radial disc extension in our context. Finally, $e_{\rm pl}$ is the planet’s eccentricity.}

\item[(ii)] \rev{Self-gravitating disc: the timescale for a planet-disc system is taken from Equation 8 of \citet{Terquem&Ajmia2010}, which we have adopted for our case, as follows:}
\begin{equation} \label{appeq:t_KL}
t_{\rm ZKL} =0.42\left[{\rm sin^2}(i_{\rm pl})-\frac{2}{5}\right]^{-1/2}{\rm ln}\left(\frac{e_{\rm max}}{e_{\rm pl}}\right) \tau_{\rm ZKL},
\end{equation}
with
\begin{equation}
e_{\rm max} = \sqrt{1-\frac{5}{3}{\rm cos^2}(i_{\rm pl})},
\end{equation}
and
\begin{equation}
\tau_{\rm ZKL} = \frac{(1+n)\left(1-\eta^{2-n}\right)}{(2-n)\left(-1+\eta^{-1-n}\right)}\frac{M_*}{m_{\rm disc}}\left(\frac{R_{\rm out}}{a_{\rm pl}}\right)^3\frac{T_{\rm pl}}{2\pi},
\end{equation}
\rev{where $i_{\rm pl}$ is the initial planet inclination, $m_{\rm disc}$ the disc mass, $\eta$ is the ratio between the inner rim of the disc ($R_{\rm in}$) and the outer rim ($R_{\rm out}$), and $n$ is related power law that defines the surface density such as $\Sigma \propto r^{-n}$. For our particular case, we set $n=1$, and $\eta=0.9$.}
\end{itemize}

 \section{Inclination precession timescales}
\label{app:np}

As long as orbital eccentricities and inclinations are low, the classical Laplace--Lagrange theory provides analytical solutions to the orbital evolution of $N$ mutually perturbing bodies around a massive central object.
After the common substitution of orbital inclinations and longitudes of the ascending nodes with
\begin{equation}
  q_i \equiv I_i \cos\Omega_i \qquad \text{and} \qquad  p_i \equiv I_i \sin \Omega_i,
\end{equation}
the solution is given by
\begin{eqnarray}
  q_i (t) &=& \sum_{i = 1}^N I_{ij} \cos(f_j t + \gamma_j),\\
  p_i (t) &=& \sum_{i = 1}^N I_{ij} \sin(f_j t + \gamma_j),
\end{eqnarray}
where the $f_j$ are the frequencies of the individual terms and $I_{ij}$ the amplitudes. The frequencies are the eigenvalues of the pair-wise interaction matrix. For a belt and a planet on non-overlapping orbits interacting with each other, this matrix is given by the four entries
\begin{eqnarray}
  B\sbs{bb} &=& -\frac{n\sbs{b}}{4} \frac{m\sbs{p}}{M_* + m\sbs{b}} \alpha\sbs{bp}^2 b_{3/2}^{(1)}(\alpha\sbs{bp}),\\
  B\sbs{bp} &=& \frac{n\sbs{b}}{4} \frac{m\sbs{p}}{M_* + m\sbs{b}} \alpha\sbs{bp}^2 b_{3/2}^{(1)}(\alpha\sbs{bp}) = -B\sbs{bb},\\
  B\sbs{pp} &=& -\frac{n\sbs{p}}{4} \frac{m\sbs{b}}{M_* + m\sbs{p}} \alpha\sbs{pb}^2 b_{3/2}^{(1)}(\alpha\sbs{pb}), \\
  B\sbs{pb} &=& \frac{n\sbs{p}}{4} \frac{m\sbs{b}}{M_* + m\sbs{p}} \alpha\sbs{pb}^2 b_{3/2}^{(1)}(\alpha\sbs{pb}) = -B\sbs{pp},
\end{eqnarray}
where ``b'' stands for belt and ``p'' for planet. The ratio of semi-major axes is $\alpha\sbs{bp} = a\sbs{b}/a\sbs{p} \equiv \alpha$. The $n$ are the orbital frequencies, which are related to the orbital periods $P\sbs{b} = 2\pi/n\sbs{b}$ and $P\sbs{p} = 2\pi/n\sbs{p} = P\sbs{b} \alpha\sbs{bp}^{-3/2}$. The stellar mass is given by $M_*$, and the masses of belt and perturber are $m\sbs{b}$ and $m\sbs{p}$, respectively. The Laplace coefficient can be expressed as \citep[cf.][]{Hahn2003,Sefilian+2019}
\begin{align}
    b_{3/2}^{(1)}(\alpha)
    &= 2\times \frac{\frac{1+\alpha^2}{(1-\alpha)^2} E(\chi) ~-~ K(\chi)}{\pi \alpha (1 + \alpha)} \nonumber
    \\ &=\left\{\begin{array}{ll}
             3\alpha^{-4} + \tfrac{45}{8}\alpha^{-6} + \mathcal{O}(\alpha^{-8}), & \text{for}~\alpha \gg 1,\\
             3\alpha + \tfrac{45}{8}\alpha^{3} + \mathcal{O}(\alpha^{5}), & \text{for}~\alpha \ll 1,
           \end{array}\right.
\end{align}
where $K(\chi)$ and $E(\chi)$ are complete elliptical integrals of the first and second kind, respectively, and
\begin{equation}
  \chi \equiv \frac{2\sqrt{\alpha}}{1 + \alpha}.
\end{equation}
This expression for the Laplace coefficient is valid for all $\alpha > 0$, including $\alpha\sbs{bp} = a\sbs{b}/a\sbs{p} > 1$. Hence, no distinction between inner and outer perturbers is necessary.
Given that $\alpha\sbs{bp} = 1/\alpha\sbs{pb}$ and
\begin{equation}
  b_{3/2}^{(1)}(\alpha) = \alpha^{-3} b_{3/2}^{(1)}(1/\alpha),
\end{equation}
the diagonal elements of the interaction matrix are related as follows:
\begin{align}
  B\sbs{pp} &= B\sbs{bb} \frac{n\sbs{p}}{n\sbs{b}} \frac{m\sbs{b}}{m\sbs{p}} \frac{M_* + m\sbs{b}}{M_* + m\sbs{p}} \frac{\alpha\sbs{pb}^2}{\alpha\sbs{bp}^2} \frac{b_{3/2}^{(1)}(\alpha\sbs{pb})}{b_{3/2}^{(1)}(\alpha\sbs{bp})} \nonumber\\
  &= B\sbs{bb} \alpha\sbs{bp}^{1/2}\frac{m\sbs{b}}{m\sbs{p}} \sqrt{\frac{M_* + m\sbs{b}}{M_* + m\sbs{p}}}\nonumber\\
  &\approx B\sbs{bb} \alpha\sbs{bp}^{1/2}\frac{m\sbs{b}}{m\sbs{p}},
\end{align}
where the last simplification is based on the assumptions that $m\sbs{b} \ll M_*$ and $m\sbs{p} \ll M_*$.

The only non-zero eigenvalue of the above $2\times 2$ matrix $B$ is \citep[e.\,g.,][]{Murray&Dermott1999}
\begin{equation}
  f = B\sbs{bb} + B\sbs{pp} \approx B\sbs{bb} \left(1 + \alpha\sbs{bp}^{1/2}\frac{m\sbs{b}}{m\sbs{p}}\right),
\end{equation}
corresponding to a full inclination precession period
\begin{align}
  t\sbs{ip} &= \frac{2\pi}{|f|}
        \approx \frac{4P\sbs{b}}{\alpha\sbs{bp}^2 b_{3/2}^{(1)}(\alpha\sbs{bp})} \, \frac{M_*}{m\sbs{p} + \alpha\sbs{bp}^{1/2} m\sbs{b}}
                \nonumber\\
              &= \frac{4P\sbs{b}}{3} \frac{M_*}{m\sbs{p} + \alpha\sbs{bp}^{1/2} m\sbs{b}}
                \left[1 - \tfrac{15}{8}\alpha\sbs{bp}^2 + \tfrac{25}{32}\alpha\sbs{bp}^4 + \mathcal{O}(\alpha\sbs{bp}^6)\right]\nonumber\\ &\times 
                \left\{\begin{array}{ll}
                  \alpha\sbs{bp}^2  & \text{for}~\alpha\sbs{bp} \gg 1,\\[1mm]
                  \alpha\sbs{bp}^{-3}  & \text{for}~\alpha\sbs{bp} \ll 1,
                \end{array}\right.\label{eq:prec-full}
\end{align}
where  $\alpha\sbs{bp} > 1$ for all cases considered in this work. The precession frequency differs across an extended belt.

For a low-mass belt, where $m\sbs{b} \ll m\sbs{p}$, the corresponding differential precession period, defined as the time over which the near belt edge precesses one complete cycle further than the far one, is given by
\begin{equation}
  t_{\Delta \text{ip}} = \left(T\sbs{near}^{-1} - T\sbs{far}^{-1}\right)^{-1} = \frac{T\sbs{near}T\sbs{far}}{T\sbs{far} - T\sbs{near}} \approx \frac{t\sbs{ip}^2}{\Delta T},\label{eq:prec-diff}
\end{equation}
where
\begin{align}
  \Delta T &= \Delta a\sbs{b} \frac{\total t\sbs{ip}}{\total a\sbs{b}} 
           = \Delta \alpha\sbs{bp} \frac{\total t\sbs{ip}}{\total \alpha\sbs{bp}} \nonumber\\
           &= \Delta \alpha\sbs{bp} \left|
                    - \frac{1}{2}\frac{t\sbs{ip}}{\alpha\sbs{bp}}
                  - \frac{t\sbs{ip}}{b_{3/2}^{(1)}(\alpha\sbs{bp})} \frac{\total b_{3/2}^{(1)}(\alpha\sbs{bp})}{\total\alpha\sbs{bp}}
                \right|,
\end{align}
and hence
\begin{multline}
  t_{\Delta \text{ip}} \approx t\sbs{ip} \frac{\alpha\sbs{bp}}{\Delta\alpha\sbs{bp}}\left|
  \frac{1}{2} +\frac{\alpha\sbs{bp}}{b_{3/2}^{(1)}(\alpha\sbs{bp})} \frac{\total b_{3/2}^{(1)}(\alpha\sbs{bp})}{\total\alpha\sbs{bp}}\right|^{-1}\\
  t_{\Delta \text{ip}} = t\sbs{ip} \frac{\alpha\sbs{bp}}{\Delta\alpha\sbs{bp}}\times \\ \left\{\begin{array}{ll}
\frac{2}{7}\left[1 - \tfrac{15}{14}\alpha\sbs{bp}^{-2} + \tfrac{25}{784}\alpha\sbs{bp}^{-4} + \mathcal{O}(\alpha\sbs{bp}^{-6})\right] & \text{for}~\alpha\sbs{bp} \gg 1,\\
\frac{2}{3}\left[1 - \tfrac{5}{2}\alpha\sbs{bp}^{2} +  \tfrac{175}{48}\alpha\sbs{bp}^{4} + \mathcal{O}(\alpha\sbs{bp}^6)\right]& \text{for}~\alpha\sbs{bp} \ll 1.
           \end{array}\right.\label{eq:prec-diff-expansion}
\end{multline}
This approximation provides higher-order accuracy only in the belt position $a\sbs{b}$, \textit{not} in the relative belt width $\Delta a\sbs{b}/a\sbs{b} = \Delta \alpha\sbs{bp}/\alpha\sbs{bp}$. In addition, this result is only valid for $m\sbs{b} \ll m\sbs{p} \ll M_* $ as well as low eccentricities and inclinations.

\section{Settling time and Stokes number}
\label{app:tau and St}

The vertical motion for a particle in a non-self-gravitating disc composed of gas and dust is given by:
\begin{equation}\label{appeq:EVM}
\ddot{z} = - \left(\Omega_k{\rm St^{-1}}\right)\dot{z} -\Omega_k^2 z,
\end{equation}
where $\Omega_k$ is the Keplerian frequency, and St is the Stokes number. The type of solutions of this linear ordinary differential equation will depend on the discriminant of its associated characteristic equation. The discriminant of the characteristic equation is:
\begin{equation}
 \Delta = \left(\Omega_k{\rm St^{-1}}\right)^2 -4\Omega_k^2.
\end{equation}
Our case of interest is when $\Delta < 0$, which corresponds to $\mathrm{St} > 1/2$ (although in the literature, this condition is also expressed as $\mathrm{St} > 1$). This value for the Stokes number leads to a solution for the Equation \ref{appeq:EVM} as:
\begin{equation}
z(t) = z_o e^{-t/\tau}\mathrm{cos}\left( \omega t + \phi_o\right),
\end{equation} 
where $z_o$ and $\phi_o$ are constants of integration, and $\tau$ is the settling time given by 
\begin{equation}\label{appeq:tau}
\tau = 2\ \mathrm{St} \Omega_k^{-1},
\end{equation}
and
\begin{equation}
\omega = \frac{\Omega_k}{2\ \rm St} \sqrt{4\ \mathrm{St}^2-1}.
\end{equation}

The Stokes number is computed using equation 4 of \citet{Birnstiel+2016}:
\begin{equation}\label{appeq:St0}
{\rm St} = \frac{\pi}{2} \frac{\rho_s s}{\Sigma_g},
\end{equation}
where $\rho_s$ is the intrinsic dust grain density, $s$ is the grain size, and $\Sigma_g$ is the gas surface density. \rev{Combining Equations \ref{appeq:tau} and \ref{appeq:St0}, we obtain}
\begin{multline}
    \tau = 1.873\ \left[\mathrm{yr}\right]\times \\  \left( \frac{\rho_s}{\rm g/cm^3} \right) \left( \frac{s}{\rm m} \right) \left( \frac{\Sigma_g}{ \rm  M_\oplus/au^2}\right)^{-1} \left( \frac{M_*}{ M_\odot}  \right)^{-\frac{1}{2}} \left( \frac{r}{\rm au}\right)^{\frac{3}{2}},
\end{multline}
where $M_*$ is the mass of the star, and $r$ is the radial distance.

Adopting a disc with a central cavity and assuming that the gaseous component is characterised by a constant surface density, we can rewrite the gas surface density in Equation \ref{appeq:St0} as
\begin{equation}\label{appeq:St}
{\rm St} = \frac{\pi^2}{2} \frac{\rho_s s}{M_g} \Delta R,
\end{equation}
where $M_g$ is the total mass of gas, while $\Delta R = r_{\rm out}^2 - r_{\rm in}^2$: the difference between the squares of the outer rim and the inner rim of the disc. For the values of $r_{\rm in}=100$ au, $r_{\rm out}=110$ au,  and $\rho_s=3$ g/cm$^3$ we obtain the grain size in which the Stokes number is larger than 1 as a function of the total gas mass as following:
\begin{equation}
s > 8.584\cdot10^{-6} \left[\rm m\right] \left( \frac{M_g}{M_\oplus}\right).
\end{equation}
Then for the total gas mass values of $10^{-3}\ M_\oplus$ and $10^{1}\ M_\oplus$ we find the lowest limit for the grain size of $\sim8.58$ nm and $\sim85.8$ $\micro$m, respectively. In this case, the settling time will be proportional to the Stokes number.

Alternatively, and using Equations \ref{appeq:tau} and \ref{appeq:St}, we can obtain the settling time as a function of the disc parameters,
\begin{multline}
    \tau = 5.885\ \left[\mathrm{yr}\right]\times \\  \left( \frac{\rho_s}{\rm g/cm^3} \right) \left( \frac{s}{\rm m} \right) \left( \frac{\Delta R}{\rm au^2} \right) \left( \frac{M_*}{ M_\odot}  \right)^{-\frac{1}{2}} \left( \frac{M_g}{ M_\oplus}\right)^{-1} \left( \frac{r}{\rm au}\right)^{\frac{3}{2}}.
\end{multline}

\label{lastpage}
\end{document}